\documentstyle[12pt,epsfig,amsmath,amssymb]{report}
\title{Coherent transverse instability of bunched beams in circular accelerators}
\author{V. Balbekov }
\date{November 2020}

\begin{document}

\maketitle

\noindent
{\bf CONTENT}
\\
\\
{\bf 1.~Basic relations}
\\
\\
{1.1.~Definitions and designations.}
\\
\\
{1.2.~Equation of coherent betatron oscillations}
{\footnotesize\sl(smooth linear betatron oscillations; nonlinear space charge field; wake field; chromaticity; coherent oscillations; Laplace transform)}.
\\
\\
{1.3.~Landau damping of a coasting beam}{\footnotesize\sl~(restricted or unrestricted momentum spread?)}.
\\
\\
{\bf 2.~Single bunch general equations} {\footnotesize\sl(short wake)}.
\\
\\
{\bf 3.~Hollow bunch in a square well} 
\\
\\
3.1.~The bunch equation
{\footnotesize\sl(phase as a longitudinal coordinate)}.
\\
\\
3.2.~Hollow bunch eigenmodes
{\footnotesize\sl(two clusters of the eigentunes)}.
\\
\\
3.3.~Approximation of separated multipoles
{\footnotesize\sl(head-tail instability with space charge)}.
\\
\\
3.4.~Solution by an expansion
{\footnotesize\sl(truncated series of equations with wake field; 
the tune coalescence vs SC tune shift; applicability)}.
\\
\\
3.5.~TMCI: numerical solution
{\footnotesize\sl (no expansion used; arbitrary SC tune shift)}.
\\
\\
3.6.~Synchrotron tune spread
{\footnotesize\sl (like the coasting beam)}.
\\
\\
{\bf 4.~Square bunch instability (parabolic well; square bunch).}
\\
\\
4.1.~The bunch equation{\footnotesize\sl~(linear synchrotron oscillations; integral equation)}.
\\
\\
4.2.~The bunch eigenmodes
{\footnotesize\sl(Legendre functions; addition theorem; the eigentune equation; 
two eigentune clusters; ultimate SC tune shift)}.
\\
\\
4.3.~Solution by an expansion {\footnotesize\sl (wake field; infinite series of equations)}.
\\

$\quad$4.3.1~Head-tail instability 
{\footnotesize\sl (separated modes; head-tail instability)}.

$\quad$4.3.2~TMCI: multimode solution
{\footnotesize\sl (truncated series, threshold, applicability)}.
\\
\\
4.4.~TMCI: numerical solution
{\footnotesize\sl (no expansion, arbitrary SC tune shift)}.
\\
\\
{\bf 5.~Arbitrary bunch shape (parabolic well, flat wake).}
\\
\\
5.1.~The bunch integral equation.
\begin{quote}
{\footnotesize\sl 
An integral equation of the coherent oscillations is obtained for arbitrary bunch 
with linear synchrotron oscillations.
The space charge, wake field, and chromaticity are taken into account.
The square bunch (Ch.~4) is a partial case.
}
\end{quote}
5.2.~Are the coherent oscillations damping at a large space charge?
\begin{quote}
{\footnotesize\sl 
Because the Laplace transform used, the equation directly relates to a half of the complex frequency plane, and an analytical continuation is needed.
Like the square bunch, the eigentunes are separated in 2 parts with the Landau damping in the lower cluster, and with undumped oscillations in the upper one.
}
\end{quote}
5.3.~Numerical solution (the upper cluster).
\begin{quote}
{\footnotesize\sl
A model for the upper cluster is proposed in a form of the second order differential equation. 
The case of a flat wake is considered at zero chromaticity. 
The TMCI threshold tends to 0 at the SC increasing, at any bunch shape.
}
\end{quote}


\chapter{Basic relations}



\section{Definitions and designations}


A beam consisting of $J$~bunches uniformly distributed along the azimuth of a ring accelerator of radius $R$ on the interval of $\Delta\theta=2\pi/J$ is the main subject of the paper.
It is assumed that any bunch has the azimuthal extension of $\pm\theta_0$ 
with $\theta_0\le\Delta\theta/2$ (no overlapping). 
However, it is not presumed that the bunches are identical in shape and population; in particular, some of them may be actually lacking (zero population). 

In parallel with the laboratory frame, marked by the sub-index $_L$, the rest frame will be used father, with following relation of the longitudinal coordinates:
\begin{equation}
\theta_L = \theta+\Omega_0 t 
\end{equation}
where $\Omega_0$~is the angular velocity of the beam, and $t$~is time.
Besides, the normalized local (intra-bunch) longitudinal coordinate $\tau$~will be used as well being determined for $j$-th bunch by the relation:
\begin{equation}
\theta = j\Delta\theta+\theta_0\tau,\qquad\quad |\tau|\le 1.
\end{equation}
Side by side with these Cartesian coordinates, amplitude and phase of synchrotron oscillations will be applied, for example
\begin{equation} 
 \tau = A\cos\phi,\qquad u =-A\sin\phi,\qquad \phi=\Omega_0  Q_s t
\end{equation}                         
with $Q_s$~as a tune of the synchrotron oscillations.
As a rule, the unified symbols will be applied for different presentations of any variable, with appropriate indexes added if needed. 
For example, the horizontal displacement of the beam center at azimuth $\theta_L$ can be presented in one of the following forms:
\begin{eqnarray}
\bar X_L(t,\theta_L)=\bar X_L(t,\theta+\Omega_0t)
\equiv \bar X(t,\theta)=\bar X(t,j\Delta\theta+\theta_0\tau)
\equiv \bar X_j(t,\tau)
\end{eqnarray}          
Correspondingly, the relation of convective and usual derivatives 
with respect to time is
\begin{eqnarray}
\frac{d}{dt} = \frac{\partial}{\partial t} + \Omega(p)\frac{\partial}{\partial\theta_L} + \dot p(\theta)\frac{\partial}{\partial p} = \frac{\partial}{\partial t} + [\Omega(p)-\Omega_0]\frac{\partial}{\partial\theta} +
\dot p(\theta)\frac{\partial}{\partial p} \nonumber \\ 
= \frac{\partial}{\partial t} + \dot\tau(u)\frac{\partial}{\partial\tau} + \dot u(\tau) \frac{\partial}{\partial u} = \frac{\partial}{\partial t} + \Omega_0 Q_s(A)\frac{\partial}{\partial \phi} 
\end{eqnarray}


\section{Equation of coherent \\ betatron oscillations.}


We will treat the betatron oscillations of a single particle in frame of the smooth linear model with a known betatron tune $Q_p$ which does not depend on the betatron amplitude but can depend on the longitudinal momentum.
Then, with the wake field $G$~and the space charge field $E$~being  taken into account, equation of betatron oscillations of the particle in the rest frame is
\begin{equation}
\frac{d^2x}{dt^2}+\Omega_p^2 Q_p^2 x = 
\frac{eE\Big(\theta,x-\bar X(t,\theta),y\Big)}{m\gamma^3} +\frac{eG(t,\theta)}{m\gamma}
\end{equation}
where $x$ and $y$ are transverse coordinates of the particle, $\gamma$ is its normalized energy, $\Omega_p$ is the momentum-dependent angular velocity.
It is presumed that the intrinsic field of the beam $E$~depends on transverse deviation of the particle with respect to the beam center, position of which is denoted as $\bar X(t,\theta)$. 
This field does not depend on the beam pipe size as well as on other environments of the beam, which effect is described by the wake field $G$. The last does not depend on the coordinates $x,y$ but linearly depends on $\bar X$ (it is assumed that there is no vertical displacement of the beam as a whole). 
It can be represented in the form
\begin{eqnarray}
G(t,\theta) = e\beta c\int_0^{t} W'_1(-\beta c t')\lambda(\theta+\Omega_0t')
\bar X(t-t',\theta +\Omega_0 t')\,dt'  
\end{eqnarray}                                           
where $\beta$ is the beam normalized velocity, $\lambda(\theta)$ is the beam linear density (dimension cm$^{-1}$), and $W'_1(\theta)$ is the transverse wake field potential per unit of length (dimension cm$^{-3}$). 

The next step is an averaging of Eq.~(1.6) over all particles with the longitudinal momentum $p$ at given $\theta$ and $t$.  
Let the function $X(t,\theta, p)$ is an average transverse displacement of all the particles located in some point of the longitudinal phase space $\theta,p$.
Then the partial transverse distribution of these particles is described by the function $\rho_\perp(x-X(t,\theta,p),y)$ where $\rho_\perp(x,y)$ is the normalized steady-state beam density.
Consequently, the averaging of Eq.~(1.6) results in
\begin{equation}
\frac{d^2 X}{dt^2}+\Omega_p^2Q_p^2 X = \frac{e}{m\gamma^3}
\int_{-\infty}^{\infty}E(\theta,x-\bar X,y)
\rho_\perp(x-X,y)\,dx dy + \frac{eG}{m\gamma}
\end{equation}
Relation between the values $X$ and $\bar X$ follows from their definitions  as the partial and the total beam displacement:
\begin{eqnarray}
\int X(t,\theta,p) F(\theta,p)\,dp
=\bar X(t,\theta) \int F(\theta,p)\,dp 
\end{eqnarray}
where $F$ is the beam distribution function.
Because both $X$ and $\bar X$ are presumed to be rater small in comparison with the beam diameter, it is possible to expand the sub-integral functions in Eq.~(1.8) in the Taylor series, obtaining
\begin{equation}
\frac{d^2 X}{dt^2}+\Omega_p^2 Q_p^2 X = 2\Omega_0^2Q_0\Delta Q(\theta)\,(X-\bar X)+\frac{eG}{m\gamma}              
\end{equation}
where $\Delta Q(\theta)$ is the space charge driven tune shift at azimuth $\theta$, averaged over transverse coordinates
\footnote{It is taken into account that $\rho_\perp$~and $E$~are the even and the odd functions of $x$, correspondingly.
For an elliptical beam of constant density, $\Delta Q$~is the usual incoherent tune shift.
For Gaussian beam, $\Delta Q$~is exactly a half of the tune shift 
of small betatron oscillations.}:
\begin{equation}
\Delta Q(\theta) =\frac{e}{2m\gamma^3\Omega_0^2Q_0}\int_{-\infty}^{\infty} \frac{\partial E}{\partial x} (\theta,x,y)  \rho_\perp(x,y)\,dxdy
\end{equation}
It follows from these relations that nonlinear dependence of the beam field on transverse coordinates does not manifest itself in the equations of coherent betatron oscillations, and does not affect the consequent consideration. 
However, it should be noted that a disregard of the external nonlinearity is an essential point on the way to this conclusion.

With a reasonable assumption $\Delta Q\ll Q$, Eq. (1.10) can be reduced to the first-order equation:
\begin{eqnarray}
\frac{dX}{dt}+i\,\Omega_p Q_p X \simeq i\,\Omega_0
\Delta Q(\theta)\,(X-\bar X) \hspace{15mm} \nonumber \\
+\,\frac{ie^2\beta c}{2m\gamma\Omega_0Q_0}\int_0^{t} W'_1(-\beta c t')\,
\lambda(\theta+\Omega_0 t')\,\bar X(t-t',\theta +\Omega_0 t')\,dt' \end{eqnarray}
where Eq.~(1.7) and Eq.~(1.9) are taken into account. 
Remembering that $d/dt$ is the convective derivatives with respect to time, and using Eq.~(1.5), one can represent the previous equation as
\begin{eqnarray}
\frac{\partial X}{\partial t} 
+(\Omega_p-\Omega_0)\frac{\partial X} {\partial\theta}
+\dot p(\theta)\frac{\partial X}{\partial p}
+i\,\Omega_p Q_p X-i\,\Omega_0\Delta Q(\theta)\,(X-\bar X) 
\nonumber \\
=\,\frac{ie^2\beta c}{2m\gamma\Omega_0Q_0}
\int_0^{t} W'_1(-\beta c t')\,\lambda(\theta+\Omega_0 t')\,
\bar X(t-t',\theta +\Omega_0 t')\,dt' \hspace{5mm}
\end{eqnarray}

The Laplace transform will be applied as the next step of the calculations. 
Multiplying Eq.~(1.13) by $\exp(i\omega t)$ with Im~$\omega>0$,
integrating over $t$, and using the notations
\begin{subequations}
\begin{eqnarray}
X_{\omega}(\theta,p) = \int_0^\infty X(t,\theta,p)\exp(i\omega t)\,dt,
\end{eqnarray}
\vspace{-5mm}
\begin{eqnarray}
\bar X_{\omega}(\theta)=\int_0^\infty \bar X(t,\theta)
\exp(i\omega t)\,dt \;\;
\end{eqnarray}
\end{subequations}
obtain
\begin{eqnarray}
-\!X_{in}\!-\!i(\omega\!\!-\!\!\Omega_p Q_p) X_\omega 
\!+\!(\Omega_p\!\!-\!\!\Omega_0)\frac{\partial X_\omega} {\partial\theta}
\!+\!\dot p(\theta)\frac{\partial X_\omega}{\partial p}       
\!-\!i\,\Omega_0\Delta Q(\theta)\,(X_\omega\!\!-\!\!\bar X_\omega)
\nonumber \\ 
= \,\frac{ie^2\beta c}{2m\gamma\Omega_0Q_0}\int_0^{\infty} 
W'_1(-\beta ct)\,\lambda(\theta+\Omega_0t)\,
\bar X_\omega (\theta+\Omega_0t) \exp(i\omega t)\,dt
\hspace{12mm}
\end{eqnarray}
where $X_{in}=X(0,\theta,p)$ is the initial value of the beam displacement.


\section{Landau damping of a coasting beam. }


The case of a coasting beam is shortly considered in this section,
mostly to exam some properties of the Landau damping, which are essential for subsequent analysis.

Longitudinal momentum of any particle is constant $(\dot p=0)$ in the case, the SC tune shift $\Delta Q$ does not depend on $\theta$, and $\lambda=N/2\pi R$  is constant as well, with $N$ as number of particles in the beam.
It is easy to see that the longitudinal dependence of all variables in Eq.~(1.15) is $\,\propto\!\exp(ik\theta),$ with an integer $k$. 
It results in the following expression:
\begin{eqnarray}
X_\omega(p)=\frac{(\Delta Q+\Delta Q_{Wk})\bar X_\omega+iX_{in}(p)/\Omega_0}{\hat\nu-\zeta_k (p-p_0)/p_0}
\end{eqnarray}
where
\begin{subequations}
\vspace{-5mm}
\begin{eqnarray}
\hspace{-8mm}
\hat\nu = \frac{\omega}{\Omega_0}-Q_0+\Delta Q, 
\end{eqnarray}
\vspace{-7mm}
\begin{eqnarray}
\zeta_k = \frac{p_0}{\Omega_0} \frac{d(\Omega[k+Q])}{dp},\qquad
\end{eqnarray}
\vspace{-7mm}
\begin{eqnarray}
\Delta Q_{Wk} =-\frac{r_0RN}{4\pi\beta^2\gamma Q_0}
\int_{-\infty}^0 W'_1(\beta ct)\,
\exp\left(-i\,[\omega+k\Omega_0]\,t\right) d(\beta c t)
\end{eqnarray}
\end{subequations}
with $r_0=e^2/mc^2$ as the particle electromagnet radius.
A relation of the variables $X_{\omega}(p)$ and $\bar X_\omega$ follows from Eq.~(1.9), and in this simple case can be used in the form
\begin{eqnarray}
\bar X_\omega=\int X_{\omega}(p)F(p)\,dp
\end{eqnarray}
with an appropriate normalized momentum distribution function $F(p)$.
Therefore the expression can be obtained from Eq.~(1.16): 
\begin{eqnarray}
\bar X_{\omega} = \frac{i\,\int \frac{X_{in}(p)F(p)\,dp}
{\hat\nu-\zeta_k (p-p_0)/p_0}}
{\Omega_0\left[1-(\Delta Q+\Delta Q_{Wk})\int\frac{F(p)\,dp}
{\hat\nu-\zeta_k (p-p_0)/p_0}\right]}
\end{eqnarray}
It is convenient to use the replacements: 
$\,p-p_0=s\Delta p$, and $\,F(p)dp=\Phi(s)ds,$ 
with $s$ as a new variable like momentum, and $\Delta p$ as a characteristic parameter of the momentum distribution (e.g. the total or rms spread). 
Then Eq.~(1.19) can be written in the form
\begin{eqnarray}
\bar X_{\omega} = \frac{1}{w(z)-w_k} 
\int \frac{X_i(s)\Phi(s)\,ds}{z-s} 
\end{eqnarray}
We write the appearing parameters for the case of an even function $\Phi(s)$: 
\begin{eqnarray}
 w(z) = \int \frac{\Phi(s)\,ds}{z-s},\qquad
 z = \frac{\hat\nu p_0}{|\zeta_k|\Delta p}, \qquad
 w_k =\frac{|\zeta_k|\Delta p}{p_0(\Delta Q+\Delta Q_{Wk})},
\end{eqnarray}
Specific form of the function $X_i(s)\propto X_{in}(p_0+s[p-p_0])$ does not matter further. 

The time-dependent function $\bar X(t)$ should be calculated by the inverse Laplace transform:
\begin{eqnarray}
\bar X(t) =\frac{1}{2\pi}\int_{-\infty+ic}^{\infty+ic}\bar X_\omega\exp(-i\omega t)\,d\omega
\end{eqnarray}
where the integration path runs above any of the integrand singularities. 
It is pertinently to remind here that the function $X_\omega$ is determined in the upper semi-plane Im$\,\omega>0$.
However, calculation of the above integral requires often an analytical expansion of the function in entire complex space.  
In particular, it has the poles which can be found by solution of the dispersion equation
\begin{eqnarray}
w(z)=w_k
\end{eqnarray}
Several examples are considered below.


\subsubsection{Square distribution}

The simple distribution function is considered as an example: $F(p)=1/(2\Delta p$) at $|p-p_0|<\Delta p$, that is $\Phi(s)=1/2$ at $|s|<1$.
According to Eq~(1.21) 
\begin{eqnarray}
w(z) = \frac{1}{2}\ln\frac{z+1}{z-1}
\end{eqnarray}
It is the analytical function in entire complex plane $z=x+iy$ except the branch cut $-1\le x \le 1$, and besides the function Im$\,w$ is negative/positive in the upper/lower edges of the cut.
As it follows from Eq.~(1.21) and Eq.~(1.23), the function $\bar X_\omega$ has also the pole $\,z=\coth w_k\,$ that is 
\begin{eqnarray}
\omega=\Omega_0\left[Q_0-\frac{|\zeta_k|\Delta p}{p_0}
\coth\frac{|\zeta_k|\Delta p}{(\Delta Q+\Delta Q_{Wk})p_0}\right]\quad
\end{eqnarray}
The inverse Laplace transform, Eq.~(1.22), is reducible to two separate 
contour integrals embracing the cut (i), or the pole (ii), and describing:
(i) the time evolution of the initial deviation (decoherence), or (ii) the beam coherent oscillations with eigenfrequency given by Eq.~(1.25).
Without a wake, that is at~Im$\,\Delta Q_W =0\,$ and any $\,\Delta Q$, the real coherent tune is located outside the area where the particle individual tunes are spread.
With a wake, the tune acquires an imaginary addition (possible instability) without any threshold. 
It particular, the eigentune is $\omega\simeq\Omega_0(Q_0+\Delta Q_{Wk})$ both at very small $\Delta p/p_0$ and at rather large $\Delta Q$. 


\subsubsection{Parabolic distributions}


The normalized distribution function and corresponding integral are in the case
\begin{eqnarray}
\Phi(s)=\frac{3}{4}\,(1-s^2),\qquad 
w(z)=\frac{3}{2}\left(z+\frac{1-z^2}{2}\ln\frac{z+1}{z-1}\right)
\end{eqnarray}
Like the previous case, $w(z)$ is analytical function in the complex plane $z$, excluding the segment $-1<x<1$.
Therefore, at real $w_0$, possible solutions of Eq.~(1.23) also can be only in the region $|x|>1,\,y=0$.
However, one more condition should be taken into account now because $|w(z)|\le 1.5$ in the pointed segment.
Parameter $w_0$ should satisfy similar condition which can be written in the form
$\,\alpha|w_0|<1\quad {\rm that~is}\,$ that is
\begin{eqnarray}
\Delta Q>\alpha\left|\frac{\zeta_k\Delta p}{p_0}\right|
\end{eqnarray}
where  $\alpha=2/3$.
Otherwise, Eq.~(1.23) does not have solutions, that is the beam has no coherent eigenmodes. 

The main circumstance in the considered cases is that the (normalized) distribution is restricted by the frame $|x|<1$.
Therefore many similar distributions result in Eq.~(1.27) condition, with appropriate coefficients.
For example, $\alpha=2n/(2n+1)$ for the distributions $\,\Phi(s)\propto (1-s^2)^n$ with different $n$, including considered above cases $n=0$ and $n=1$.
Therefore the "threshold" value of the space charge tune shift may be introduced as
\begin {eqnarray}
\Delta Q_{th} = \left|\frac{\zeta_k\Delta p}{p_0}\right|
\end{eqnarray}
with $\pm\Delta p$ as the total momentum spread.
The beam coherent oscillations are possible only above the threshold, that is at $\,\Delta Q>\Delta Q_{th}$.
Wherein their amplitude can increase under the influence of the wake field, that is the equation really provides the instability threshold
\footnote{It is necessary to take into account that~~Im$\,\Delta Q_{Wk}$ can be positive or negative, dependent on $k$}.
No coherent oscillations are possible under the threshold, 
and any initial perturbation results only in the beam decoherence.


\subsubsection{Gaussian distribution}

The Gaussian distribution is characterized by the normalized function 
$\Phi(s)={\exp(-s^2/2)}/\sqrt{2\pi}$.
Corresponding function $w(z)=w(x+iy)=u+iv$ analytically continued to the whole complex plane is  
\begin{subequations}
\begin{eqnarray}
\hspace{-5mm}
w(z)=\frac{1}{\sqrt{2\pi}}\int_{-\infty}^{\infty}
\frac{\exp(-s^2/2)}{z-s}\,ds \hspace{46mm} {\rm at}\; y>0;
\end{eqnarray}
\vspace{-5mm}
\begin{eqnarray}
\hspace{-5mm}  
w(x)=\frac{1}{\sqrt{2\pi}}{\,\rm p.v.}\!\!\int_{-\infty}^{\infty} \frac{\exp(-s^2/2)}{x-s}\,ds
-i\sqrt{\frac{\pi}{2}} \exp(-x^2/2) \quad {\rm at} \;y=0,
\end{eqnarray}
\vspace{-5mm}
\begin{eqnarray}
\hspace{-5mm}  
w(z)=\frac{1}{\sqrt{2\pi}}\int_{-\infty}^{\infty} \frac{\exp(-s^2/2)}{z-s}\,ds
-i\sqrt{2\pi} \exp(-z^2/2) \qquad\; {\rm at} \;y<0.
\end{eqnarray}
\end{subequations}
We need now to solve the equation $\,w(z)=w_k\,$ with this function. 
First we consider the case without wake.
Then the parameter $\,w_k$ is real value, and the same must be valid for the function $\,w(z)$.
It follows from Eq.~(1.29) that it is possible only in the case "c", that is at $\,y<0$.
In other words, only damped oscillations are inherent in the "parabolic" beam itself without a wake.
The damping occurs because of transformation of energy of the 
coherent oscillations into an incoherent form, i.e. into the beam heating.
This phenomenon is similar to the decay of electromagnetic waves in plasma (Landau damping \footnote {L.~Landau, J. Phys. USSR, 10, 25, 1946}).

The case $y\ge0$ (steady state or instability) is possible at Im$\,w_0<0$ that is Im$\,\Delta Q_{Wk}>0$.
Then the Landau damping is compensated by a power flow from the longitudinal motion transferred by the wake field. 
Note that this happens at any $\Delta Q$ that is the instability does not have a threshold in the case, in contrast with Eq.~(1.28).

The reason for the difference is that, at the limited particle tune spread (like parabolic), the coherent tune has a location outside the range of the particle tunes, so that an energy exchange is impossible between coherent and incoherent oscillations;
however, it is unavoidable at the unlimited tune distributions like Gaussian.

The last case looks not quite realistically, because the infinite "tails" will be cut off for one reason or another, in practice. 
With the often used estimation of $\Delta p$ as 5 or 6 $\sigma_p$,
Eq.~(1.28) gives a "realistic" threshold of the Gaussian beam:
\begin {eqnarray}
\Delta Q_{th} \simeq (5-6)\left|\frac{\zeta_k\sigma_p}{p_0}\right|
\end{eqnarray}


\chapter{Single bunch general equations}


A bunched beam is considered in this section at the condition that the wake field of any bunch is very short and does not reach the neighboring bunch. 
There is no the bunch interaction in the case, and any of them  can be considered separately. 
Eq.~(1.15) is applicable for this, however, some modifications are reasonable.  

Let's introduce new variables
\begin{subequations}
\begin{equation}
Y_\omega(\theta,p)=X_\omega(\theta,p)\exp\Big(i(Q_0+\zeta)\,\theta\Big) \;\;
\end{equation}
\vspace{-5mm}
\begin{equation}
\bar Y_\omega(\theta)\;\;\; = \,\bar X_\omega(\theta) \;\;\; \exp\Big(i\,(Q_0+\zeta)\,\theta\Big)\;
\end{equation}
\end{subequations}
where $\zeta$ is the normalized chromaticity
\begin{equation}
\zeta=\frac{\Omega_0Q'(p_0)}{\Omega'(p_0)}
=\frac{\xi}{1/\gamma^2-\alpha}, \end{equation}
with $\xi$ as the usual chromaticity, and $\alpha$ as the momentum compaction factor. 
Then Eq.~(1.15) obtains the form
\begin{eqnarray}
\hspace{-55mm}
(\omega\!-\!\Omega_0 Q_0) Y_\omega 
+i\Omega_0 Q_s\frac{\partial Y_\omega}{\partial\phi}  
+\,\Omega_0\Delta Q(\theta)\,(Y_\omega \!-\! \bar Y_\omega)\,= \hspace{18mm} \\ 
-\frac{e^2\beta c}{2m\gamma\Omega_0Q_0}\int_0^{\infty}\!\!\!
W'_1(-\beta ct)\,\lambda(\theta\!+\!\Omega_0t)
\bar Y_\omega (\theta\!+\!\Omega_0t) 
\exp\big(i[\omega\!-\!\Omega_0(Q_0\!\!+\!\!\zeta)] t\big)\,dt \nonumber
\end{eqnarray}
where
\begin{eqnarray}
\Omega_0 Q_s\frac{\partial Y_\omega}{\partial\phi}  
= (\Omega_p-\Omega_0)\frac{\partial Y_\omega} {\partial\theta}
+ \dot p(\theta)\frac{\partial Y_\omega}{\partial p} = \dot\theta\,\frac{\partial Y_\omega} {\partial\theta}
+ \dot p\,\frac{\partial Y_\omega}{\partial p} 
\end{eqnarray}
It follows from this that $\phi$ is the phase of the synchrotron oscillations, $\Omega_0 Q_s$ and $Q_s$ are their frequency and tune.
The initial value $\propto X_{in}$ is omitted here  because it has no effect on the bunch eigenmodes.

It is convenient to continue the investigation by using the variables $(\tau$--$u)$ or $(A$--$\phi)$ which have been introduced by Eqs.~(1.2)-(1.3).
Because the bunch is located in the region $-1<\tau<1$,
Eq.~(2.3) can be represented as
\begin{eqnarray}
\nu Y+i\,Q_s\frac{\partial Y}{\partial \phi}+\Delta Q\,(Y-\bar Y) 
\hspace{13mm}  \nonumber \\ = \;2q \int_0^{1-\tau} w(\tau') \exp(-i\,\zeta_\tau\,\tau')\,\bar Y(\tau+\tau')\,\lambda(\tau+\tau')\,d\tau'
\end{eqnarray}
where the notations are used:
\begin{eqnarray}
\nu=\frac{\omega}{\Omega_0}-Q_0,\qquad
\zeta_\tau=(\zeta-\nu)\theta_0,
\end{eqnarray}
and
\begin{eqnarray}
qw(\tau) =-\frac{r_0R^2N_b}{4\beta\gamma Q_0}W_1(-\tau R\theta_0)
\end{eqnarray}
with $\,N_b\,$ as number of particles in the bunch, and $\,\zeta_\tau\,$ as the betatron phase advance between the bunch center and its tail.
The function $w(\tau)$ describes the wake form, and the constant factor $q$ is the wake amplitude. 
As a rule, the following normalization condition will be used further to separate the mentioned parts in Eq.~(2.7)  
\begin{equation}
 \frac{1}{2}\int_0^2 w(\tau)(2-\tau)\,d\tau = 1.
\end{equation}
In particular, $w=1$ in the simplest case when the wake has a constant value in the bunch.
Other normalization conditions are in these terms:
\begin{subequations}
\begin{equation}
\int Y(\tau,u) F(\tau,u)\,du = \bar Y(\tau)\,\lambda(\tau),\quad\,
\end{equation}
\begin{equation}
\int F(\tau,u)\,du=\lambda(\tau),              
\end{equation}
\begin{equation}
\int\lambda(\tau)\,d\tau = 1.
\end{equation}
\end{subequations}

Without restricting the generality, one can accept that $u$ is an odd function of $\phi$. 
Then Eq.~(2.5) can be replaced by the equation for the even part of the function: $Y^{(+)}(\tau,u)=[Y(\tau,u)+Y(\tau,-u)]/2$:
\begin{eqnarray}
Q_s^2\frac{\partial}{\partial\phi}\left(\frac{1}{\hat\nu}\frac{\partial Y^{(+)}}{\partial \phi}\right) + \hat\nu Y^{(+)}  
- \Delta Q\,\bar Y \hspace{15mm}   \nonumber \\
= \;2q \int_0^{1-\tau} w(\tau') \exp(-i\,\zeta_\tau\,\tau')\,\bar Y(\tau+\tau')\,\lambda(\tau+\tau')\,d\tau'
\end{eqnarray}
where 
\begin{eqnarray}
\hat\nu(\tau)=\nu+\Delta Q(\tau).
\end{eqnarray}
Because $Y^+(\phi)$ is an even periodic function, Eq.~(2.10) is applicable at $0\le\phi\le\pi$ with the boundary conditions 
\begin{eqnarray}
\frac{\partial Y^{(+)}}{\partial\phi}=0\qquad{\rm at}\qquad \phi = 0\quad{\rm and}\quad\phi=\pi.
\end{eqnarray}


\chapter{Bunch in a square potential well} 


\section{The bunch equation}


The bunch in a square potential well is considered in this section. 
Equations (2.10)-(2.12) are applicable in such a case at $\Delta Q$ and $\hat\nu$ being some constants, and $\lambda=1/2$. 
Besides of this, the particle longitudinal coordinate is a piece-wise linear function of the synchrotron phase in such a well, so that the following relations are valid at $0<\phi<\pi$:
\begin{eqnarray}
\tau = 1-\frac{2\phi}{\pi},\qquad \phi=\frac{\pi}{2}\,(1-\tau).
\end{eqnarray}
Taking these into account and restricting the consideration on the case of a constant wake $(w=1)$, represent Eq.~(2.10) in the form
\begin{eqnarray}
Q_s^2\,\frac{\partial^2Y^{(+)}}{\partial\phi^2} 
\!+\! \hat\nu^2Y^{(+)} \!-\! \hat\nu\Delta Q\,\bar Y \!
= \frac{2q\hat\nu}{\pi} \exp(-i\,\zeta_\phi\phi)
\!\!\int_0^\phi\!\!\exp(i\,\zeta_\phi\phi')\,\bar Y(\phi')\,d\phi'
\end{eqnarray}
where $\zeta_\phi=\zeta_\tau\tau'_\phi=2(\zeta-\nu)\theta_0/\pi$.
The bunch distribution function $F$ depends now on the value of $|u|$ which fulfils a role of an amplitude of the synchrotron oscillations.
In particular, $F=\delta (|u|-u_0)\lambda/2$ if the hollow bunch model is used.
Then $Y^{(+)}=\bar Y$, $Q_s=Q_{s0}$ is a constant, and Eq.~(3.2) is reduced to
\begin{eqnarray}
Q_{s0}^2\,\frac{d^2 \bar Y}{d\phi^2} + \nu\hat\nu\bar Y 
= \frac{2q\hat\nu}{\pi} \exp(-i\,\zeta_\phi\phi)
\int_0^\phi \exp(i\,\zeta_\phi\phi')\,\bar Y(\phi')\,d\phi'
\end{eqnarray}
with the boundary conditions given by Eq.~(2.12) at $\,Y^+\,=\bar Y$. 


\section{Hollow bunch eigenmodes}


At $q=0$, the eigenfunctions of Eq.~(3.3) are $\bar Y_m=\cos m\phi$ where $m=0,\,1,\,2,\dots$
Corresponding eigentunes are determined by the equations:
\begin{equation}
\hat\nu^2-m^2Q_{s0}^2=\hat\nu \Delta Q   \qquad{\rm i.e.}\qquad
\hat \nu_{\pm m}=\frac{\Delta Q}{2}\pm\sqrt{m^2Q_{s0}^2 +\frac{\Delta Q^2}{4}}.
\end{equation}
They are plotted in the left-hand panel of Fig.~3.1 against the SC tune shift at different $m$, whereas corresponding values of 
$\nu_{\pm m}=\hat\nu_{\pm m}-\Delta Q$ are plotted in the right-hand panel.
It is seen that, at $\Delta Q=0$, all the eigentunes are $\nu_{\pm m}=\pm mQ_s$. 
Corresponding eigenfunctions are proportional to $\exp(\pm im\phi)$ as it follows straight from Eq.~(2.1) at $\Delta Q=0$ and $q=0$.
At larger $\Delta Q$, the bunch spectrum is separated in two parts in accordance with the root sign.
The ultimate value of the tunes at $\,\Delta Q/Q_{s0}\rightarrow\infty$ are in the top and the bottom parts:
\vspace{-5mm}
\begin{eqnarray}
\nonumber \\
\hat\nu_{\pm m} \rightarrow \bigg\{ {\Delta Q+m^2 Q_{s0}^2/\Delta Q 
\atop -m^2 Q_{s0}^2/\Delta Q},
\qquad
\nu_{\pm m} \rightarrow \bigg\{ {m^2 Q_{s0}^2/\Delta Q 
\atop -\Delta Q-m^2 Q_{s0}^2/\Delta Q}.
\end{eqnarray} 

\begin{figure}[b!]
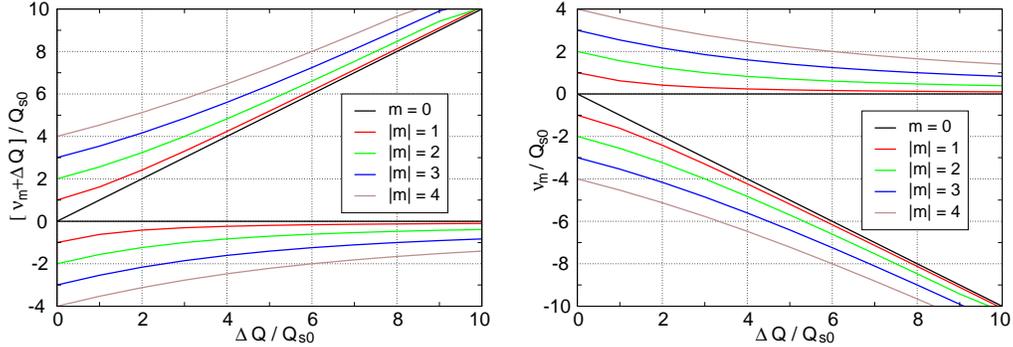

\begin{minipage}{0.47\linewidth}
\centerline{\epsfig{file=3_1a_tune.eps,width=\linewidth}}
\end{minipage}
\hspace{2mm}
\begin{minipage}{0.47\linewidth}
\centerline{\epsfig{file=3_1b_tune.eps,width=\linewidth}}
\end{minipage}
\caption{Hollow bunch in a square potential well. Different views of the bunch eigentunes are given being plotted against the SC tune shift referred to the synchrotron tune $Q_{s0}$. 
Several lowest multipoles are shown.}
\end{figure}


\section{Approximation of separated multipoles.}


 The bunch eigenfunction $\bar Y_m=\cos m\phi$ can be used as the approximate solution of Eq.~(3.3) at $|q|\ll Q_{s0}$.
 Then the tunes satisfy the equation
\begin{eqnarray}
(\nu\hat\nu-m^2Q_{s0}^2) \simeq q\hat\nu R_{m,m},\qquad m=0,\,1,\,2\,\dots
\end{eqnarray}
where
 \begin{eqnarray}
 R_{m,m}\!=\!\frac{4}{\pi^2(1\!+\!\delta_{m,0})}
 \int_0^\pi \!\!\exp(-i\zeta_\phi\phi)\cos m\phi\, d\phi \int_0^\phi \!\!\exp(i\zeta_\phi\phi')\cos m'\phi' \,d\phi'
 \nonumber \\ 
=\, \frac{2i\zeta_\phi}{\pi( m^2-\zeta_\phi^2)}
\left\{1+\frac{i\zeta_\phi(2-\delta_{m,0})}{\pi(m^2-\zeta_\phi^2)} \Big[(-1)^m \exp(-i\pi\zeta_\phi)-1\Big]\right\} \hspace{15mm}
\end{eqnarray}
Real and imaginary parts of these values are plotted in Fig.~3.2 against the parameter $\zeta_\phi$ at different $m$.
Note that these values are even and odd functions of the argument, correspondingly.
\begin{figure}[b!]
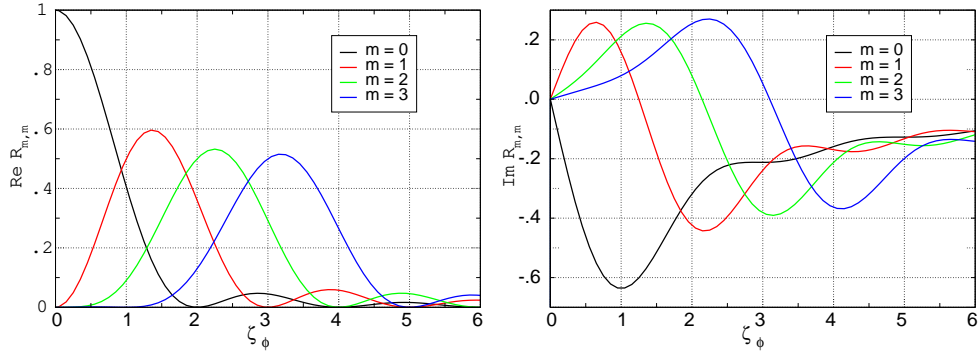

\hspace{2mm}
\begin{minipage}[h!]{0.47\linewidth}
\centerline{\epsfig{file=3_2_real.eps,width=\linewidth}}
\end{minipage}
\begin{minipage}[t!]{0.47\linewidth}
\centerline{\epsfig{file=3_2_imag.eps,width=\linewidth}}
\end{minipage}
\caption{Normalized tunes of different head-tail modes of the square bunch with a flat wake vs chromatic phase advance referred to $\pi$. 
The real part (left panel), and the imaginary one (right panel) are even/odd functions of $\zeta_\phi$,~correspondingly.}
\end{figure}
Solutions of Eq~(3.6) are close to the bunch eigentunes, given by Eq.~(3.4) and Fig.~3.1, with a small additions:  
\begin{eqnarray}
\delta\nu_{\pm m} = \frac{qR_{m,m}}{2}
\left(1\pm\frac{\Delta Q}{\sqrt{\Delta Q^2+4m^2Q_{s0}^2}}  \right)
\end{eqnarray}
It follows from Eq.~(3.8) and Fig.~(3.2) that $\,q\,$ is the tune shift of the lowest (rigid) bunch mode at $\,\zeta_\phi=0$: $\delta\nu_{+0}=qR_{0,0}=q$.  
If $\,\zeta_\phi\ne 0$, the addition to the tune includes an imaginary part. 
If the part is positive, the bunch oscillations are unstable.
It happens due to a phase advance of betatron oscillations between the bunch head and its tail ("head-tail instability"~\footnote{C. Pellegrini, Nuovo Cimento A64, 447 (1969)}).
The inequality Im$\,R_{m,m}\!>\!0$ is a condition of the bunch instability at any ${\Delta} Q$.
However, the SC tune shift essentially affects the instability rate increasing or decreasing it dependent on the root sign.
In particular, at $\Delta Q\gg Q_s$ the effects are very different in the "top" and in the "bottom" parts of the bunch spectrum being, ultimately: 
\begin{eqnarray}
\delta\nu_{\pm m} \rightarrow qR_{m,m} \times \bigg\{ 
{1-m^2 Q_{s0}^2/\Delta Q^2\rightarrow qR_{m,m}  \atop \!\!
m^2 Q_{s0}^2/\Delta Q^2 \rightarrow 0}
\end{eqnarray}

Note that the dependence of the parameter $\,\zeta_\phi$ on the tune $\nu$, following by Eq.~(2.6), is a factor of small importance in the case.
Really, in accordance with Eqs.~(3.5) and (3.8), 
$|\nu|\sim {\rm max} \{ Q_{s0},\Delta Q\}$ in order of value.
It means that corresponding part of the parameter $|\delta\zeta_\phi| \sim |\theta_0\nu|$ is typically small 
(remain that $2\theta_0$ is the bunch azimuth extent). 


\section{Solution by an expansion}


The case $\zeta_\phi=0$ is considered in this section.
Using the expansion
\begin{eqnarray}
\bar Y = \sum_{m=0}^\infty \bar Y_m\cos m\phi
\end{eqnarray}
one can transform Eq.~(3.3) to the relation 
\begin{eqnarray}
\sum_{m'=0}^\infty (\nu\hat\nu-m'^2 Q_{s0}^2)\,\bar Y_{m'} \cos m'\phi
= \frac{2q\hat\nu}{\pi} \left(\bar Y_0\phi + 
\sum_{m'=1}^\infty \bar Y_{m'} \frac{\sin m'\phi}{m'} \right).
\end{eqnarray}
Multiplying this expression by $\,\cos m\phi\,$ and integrating in the domain  $\,\phi$ from 0 to $\pi$, one can get the series of equations for the coefficients $\,\bar Y_m$
\begin{eqnarray}
(\nu\hat\nu-m^2 Q_{s0}^2)\bar Y_m = q\hat\nu \sum_{m'=0}^\infty R_{m,m'}\bar Y_{m'}
\end{eqnarray}
with the matrix
 \begin{eqnarray}
 R_{m,m}=\delta_{m,0},\qquad
 R_{m,m'\ne m}=\frac{4\big[(-1)^{m+m'}-1\big]} {\pi^2(1+\delta_{m,0})(m^2-m'^2)}.
\end{eqnarray}
Its small fragment is represented in Table 3.1.
\begin{table}[t!]
\vspace*{-5mm}
\begin{center}
\caption{Fragment of matrix $R_{m,m'}\;(A=8/\pi^2,$ flat wake).}
\vspace{3mm}
\begin{tabular}{|c|c|c|c|c|c|c|}
\hline 
$m'\rightarrow$  &~~~0~~~&~~~1~~~&~~~2~~~&~~~3~~~&~~~4~~~&~~~5~~~\\
\hline
$  ~ m=0 ~   $  &    1    &   $A/2$ &   0     & $ A/18$&    0    &  $ A/50$ \\
$  ~ m=1 ~   $  &  $ -A  $&    0    &  $ A/3$ &   0    & $ A/15$ &  0       \\
$  ~ m=2 ~   $  &     0   & $-A/3$  &   0     & $ A/5$ &    0    &  $ A/21$ \\
$  ~ m=3 ~   $  & $  -A/9$&    0    & $ -A/5$ &   0    & $ A/7$  &  0       \\
$  ~ m=4 ~   $  &    0    &$-A/15 $ &    0    & $-A/7 $&    0 &  $ A/9$     \\
$  ~ m=5 ~   $  & $ -A/25$&    0    & $ -A/21$&   0    & $-A/9$ &  0        \\
\hline
\end{tabular}
\end{center}
\vspace{-6mm}
\end{table}
We will resolve the series, truncating it by the assumption $\,\bar Y_m=0\,$ at $\,m>M$.
It is assumed that a comparison of the results with different $M$  will allow to check the correctness of the assumption.

Consequently, it is required to solve the equation $\,\det T(\nu)=0\,$ with the matrix
\begin{equation}
T_{m,m'} (\nu) = q \hat\nu R_{m,m'}+
(m^2 Q_{s0}^2-\nu\hat\nu)\,\delta_{m,m'} \qquad
m,\,m' = 0,\,1,\,2\,\dots\,M
\end{equation}
In doing so, it is necessary to take into account that the bunch tunes have to be real values at sufficiently small wake, because $\,\zeta_\phi=0$ in the case.
An appearance of complex tunes could be expected at rather large wake making known that the instability threshold is attained. 
It is important that at least two real tunes of the dispersion equations, for example $\nu_1$ and $\nu_2$, have to merge before the threshold turning these real numbers into the complex conjugated pair.
If it happens at $q=q_{\rm th},\;\nu_1=\nu_2=\nu_{\rm th}$,
the dispersion equation should have the form in a small vicinity of this point: 
\begin{eqnarray}
(\nu-\nu_{\rm th})^2={\rm const}\times (q-q_{\rm th}).
\end{eqnarray}
It is seen that the mentioned real eigentunes have to satisfy the conditions 
\begin{equation}
\frac{d\nu_{1,2}}{dq}\rightarrow\pm\infty \qquad{\rm at}\qquad q\rightarrow q_{th}.
\end{equation} 
It means that the instability can occur in the point where 
two neighboring real eigentunes coalesce    
(transverse mode coupling instability \footnote {R. Kohaupt, in Proceeding of the XI International Conference on High Energy Accelerators, p. 562, Geneva (1980).})
\\

Note that the truncated dispersion equation $\,{\rm det}\,T(\nu)=0\,$ is actually an algebraic equation of power $\,N=2M+1$.
It has $\,N\,$ roots which are real numbers at rather small $\,q$.
Therefore, the following steps can be used to resolve the problem,
and to find the TMCI threshold at arbitrary $\,\Delta Q$ and $q$:

1.~Select some values  $\,M\,$ and  $\,\Delta Q/Q_{s0}$.

2.~Take some trial value of $\,q/Q_{s0}$.

3.~Calculate the matrix $\,T(\nu/Q_{s0})\,$ and its determinant with the chosen  parameters and variable $\,\nu/Q_{s0}$.

4.~Define amount of real roots of the equation by count how many times the determinant changes sign at an increase of $\,\nu$.

5.~Repeat the shot with a higher value of $\,|q/Q_{s0}|$ until the number of the real roots decreases.
It will mean that a pair of complex roots appear and that the reached value of $\,q$ is just the TMCI threshold with taken SC tune shift, at given approximation.

6.~Check the convergence of the results by comparison of the thresholds with different $\,M$.
\\

The result very strongly depends on the wake sign. 
At $\,q>0$, the calculated threshold almost does not depend on the number $M$ and appears as a monotone decreasing function of $\Delta Q$ located in the tight limits 
\begin{eqnarray}
\frac{0.57}{1+1.21\Delta Q/Q_{s0}}< \frac{q}{Q_{s0}} 
<\frac{0.57}{\sqrt{1+1.46\Delta Q^2/Q_{s0}^2}}
\end{eqnarray}
The results with negative wake $\,q<0$ are represented in Fig.~3.3 where the TMCI threshold is plotted against the SC tune shift at different $\,M$.
The black drop-down curve belongs to all the approximations.
It is seen that the trace of any higher approximation follows the course which the lower approximations have charted, and provides a continuation of the line to the higher $\,\Delta Q$.
As a result, almost linear dependence occurs at $\Delta Q/Q_{s0}>6$:
\begin{equation}
 q \simeq -\frac{7Q_{s0}+5\Delta Q}{4}.   
\end{equation}
However, the coming back lines of different $M$ do not confirm each other, that is they cannot be accepted as credible results. 
On the contrary, the kink of the line means the end of applicability of the given approximation.
\begin{figure}[h!]
\vspace{11mm}
\centerline{\epsfig{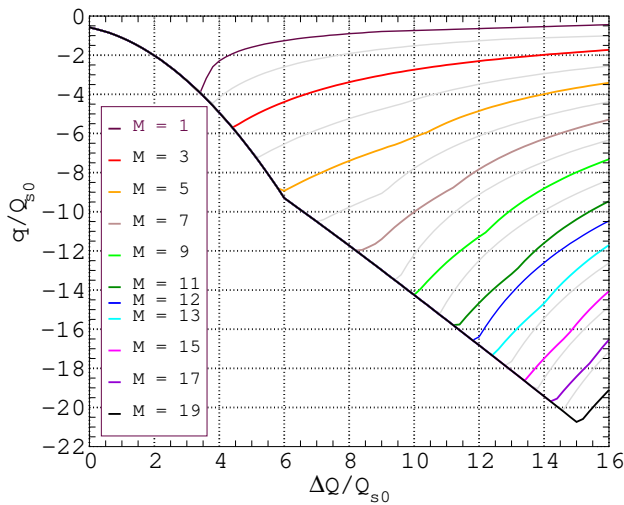}}
 \caption{TMCI threshold of a hollow bunch with a flat wake in a square potential well against 
 SC tune shift.
Different curves are obtained with different $\,M = 1,\,...\,19$.
Each curve has a restricted region of applicability which becomes
wider when $\,M\,$ is more.
The drop-down parts of the curves merge forming the sole black line.
The rising lines do not confirm each other marking ends of the applicability regions at given $\,M$.}
\end{figure}


\section{TMCI: numerical solution.}

 
It follows from the previous sections that a rather large number of the multipoles should be taken into account to obtain an acceptable precision 
when the expansion technique is used at large value of the SC tune shift. 
The method which is considered in this section is based on a numerical solution without the expansion, and therefore it is applicable at any $\Delta Q$.
\\

We will use Eq.~(3.3) with boundary conditions (2.12) at $\zeta_\phi=0$, rewriting it in the form
\begin{equation}
\bar Y''(\phi)+{\cal P}\bar Y(\phi) = \frac{2{\cal Q}}{\pi}
\int_0^\phi \bar Y(\phi')\,d\phi',
\end{equation}
where
\begin{eqnarray}
 {\cal P}=\frac{\hat\nu(\hat\nu-\Delta Q)}{Q_{s0}^2}, 
 \qquad {\cal Q}=\frac{q\,\hat\nu}{Q_{s0}^2}.
\end{eqnarray}
With any ${\cal Q}$, the equation has a countless number of solutions which are the eigenfunctions of the system $\,\bar Y_m(\phi)$ with the eigennumbers $\,{\cal P}_m$ where $m=0,\,1,...\,$.
In this part, all solutions of Eq.~(3.19) with real ${\cal Q}$ and ${\cal P}$ will be calculated numerically by going step-by-step from the point $\phi=0$ to the point  $\phi=\pi$ at initial conditions $\,\bar Y(0)=1,\;\bar Y'(0)=0\,$ as it is required by Eq.~(2.12).
At given ${\cal Q}$, the eigenvalues ${\cal P}_m$ will be chosen as these which provide the result $\bar Y'(\pi)=0$.
The procedure is repeated so many times to obtain rather detailed functions  ${\cal P}_m({\cal Q})$.
After this, the bunch eigentunes can be calculated by using the equation
\begin{equation}
\hat\nu_{\pm m} = \frac{\Delta Q}{2}\pm\sqrt{\frac{\Delta Q^2}{4} 
+{\cal P}_m Q_{s0}^2}  \qquad {\rm at} \qquad
q = \frac{Q_{s0}^2{\cal Q}}{\hat\nu_{\pm m}}  
\end{equation}
as it follows from Eq.~(3.20).
Another convenient form of this expression is
\begin{equation}
\nu_{\pm m} =-\frac{\Delta Q}{2}\pm\sqrt{\frac{\Delta Q^2}{4} 
+{\cal P}_m Q_{s0}^2}  \qquad {\rm at} \qquad
q=\frac{\nu_{\pm m} {\cal Q}}{{\cal P}_m}  
\end{equation}

The results are represented in Fig.~3.4 where 6 lowest real eigennumbers ${\cal P}_m$ are plotted against the variable $\cal Q$.
At $\cal Q\!=\,$0, the values ${\cal P}_m$ are: $0,\,1,\,2^2,\,3^2$, etc., which fact allows to identify $m$ as the bunch eigenmode number determined by Eq.~(3.4).
The lines form loops in the $({\cal Q\!-\!P})$ plain by a coalescence of the modes: $m=$ (0-1), (2-3), (4-5), etc., which are shown by different colors in the figure.
 \begin{figure}[t!]
 \centerline{\epsfig{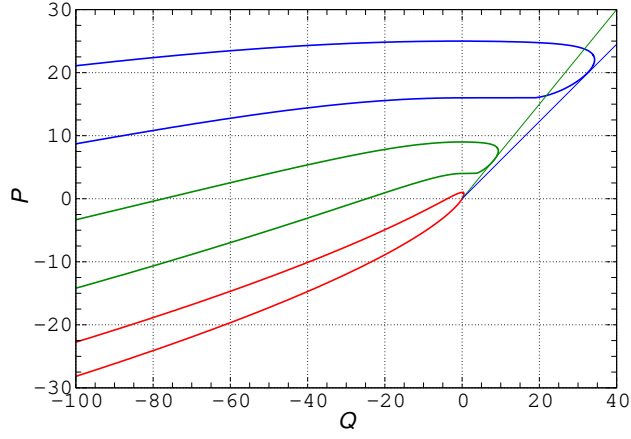}}
 \vspace{3mm}
 \caption{Lowest real eigennumbers of Eq.~(3.19).
Thin straight lines are tangent to the curves of corresponding color. They are used to determine the TMCI threshold at $\Delta Q/Q_{s0}\gg1$
(see explanation in the text and Eq.~(3.27)).}
\end{figure}

Being projected in the plane $\,(q\!-\!\nu)\,$ with help of Eq.~(3.22), each loop generates a series of curves with different $\Delta Q$ as it is shown in Fig.~3.5~a-c. 
Any point of any of these curves is a tune of some head-tail mode of the bunch.
There are turning points in the curves where the condition $dq/d\nu=0$ is fulfilled. 
It means that 2 real eigentunes coalesce in this point giving start to 2 complex conjugated tunes, that is to the TMCI instability.
The location of these points just allows to recognize dependence of the instability thresholds $q_{th}$ on $\Delta Q$.
Correspondent TMCI modes will be labeled further by the symbols $M_{m_1,m_2}$ with the thresholds $q_{m_1,m_2}$, where the multipole signs $m_i$ can be positive or negative. 
\\
\begin{figure}[t!]
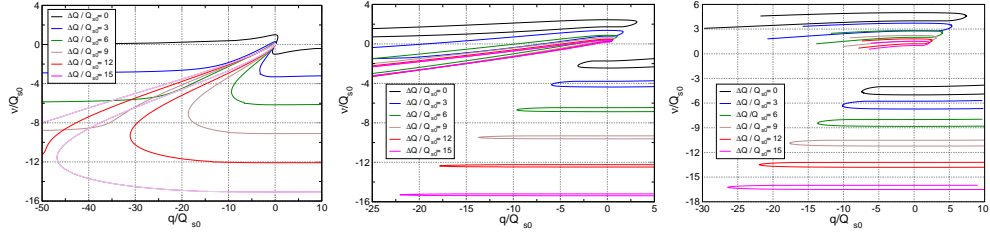

\vspace{6mm}
\hspace{1mm}
\begin{minipage}[h!]{0.31\linewidth}
\centerline{\epsfig{file=3_5a_01.eps,width=1.\linewidth}}
\end{minipage}
\begin{minipage}[t!]{0.31\linewidth}
\centerline{\epsfig{file=3_5b_23.eps,width=1.\linewidth}}
\end{minipage}
\begin{minipage}[t!]{0.31\linewidth}
\centerline{\epsfig{file=3_5c_45.eps,width=1.\linewidth}}
\end{minipage}
\vspace{3mm}
\caption{Tunes of the bunch head-tail modes against the wake strength at different SC tune shifts. 
The left hand panel refers to the modes $m=(0-1)$, middle $m=(2-3)$, right $m=(4-5)$.
The turning points where $dq=0$ mark the beginning of an instability region.}
\end{figure}

It is seen from Fig.~3.5 that the dependence of the threshold on $q$ is very sensitive to sign of the wake.  
At $q>0$, thresholds of these modes are
\begin{eqnarray}
\hspace{-5mm}
q_{0,1}\simeq\!\frac{0.57Q_{s0}}{\sqrt{1\!\!+\!\!1.5\Delta Q^2\!/Q_{s0}^2}},\;\;
q_{2,3}\simeq\!\frac{3.46Q_{s0}}{\sqrt{1\!\!+\!\!\Delta Q^2\!/Q_{s0}^2}},\;\;
q_{4,5}\simeq\!\frac{7.37Q_{s0}}{\sqrt{1\!\!+\!\!\Delta Q^2\!/Q_{s0}^2}}.
\end{eqnarray}
First of the estimations is in agreement with Eq.~(3.17) obtained by the expansion method.
Other modes have higher thresholds at any value of $\Delta Q$.

\begin{figure}[b!]
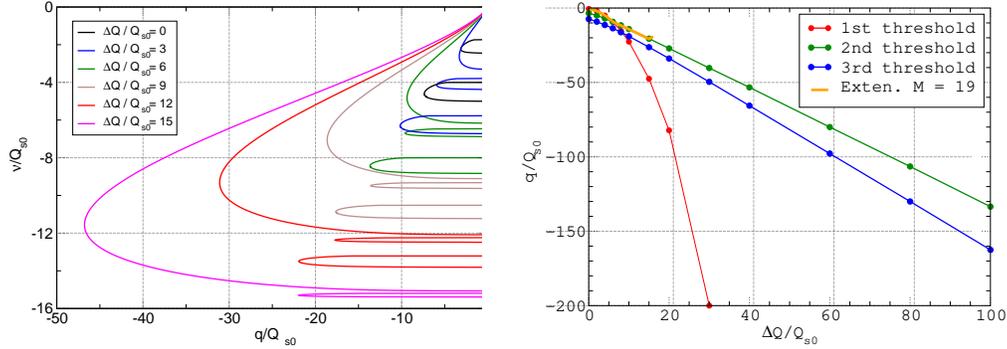

\hspace{1mm}
\begin{minipage}[h!]{0.47\linewidth}
\centerline{\epsfig{file=3_6a_abc.eps,width=\linewidth}}
\end{minipage}
\hspace{1mm}
\begin{minipage}[t!]{0.47\linewidth}
\centerline{\epsfig{file=3_6b_porog.eps,width=\linewidth}}
\end{minipage}
\caption{TMCI of the hollow bunch in a square potential well with a negative wake.
The left panel: tunes of the bunch against the wake strength at different space charge tune shifts. 
The tunes of the modes $m=(0-1)$, $m=(2-3)$, and $m=(4-5)$ are plotted at $q<0$. 
The higher mode has a higher absolute value tune shift.
The turning point of each loop indicates the TMCI threshold of the corresponding coupling mode.
The right panel: the TMCI thresholds of different modes against the space charge tune shift. 
The brown, green, and blue lines refer to the TMCI modes $M_{0,-1},\, M_{-2,-3},\,M_{-4,-5}$.
The orange-colored piece is obtained by the expansion method, Sec.~3.4.}
\end{figure}

Thresholds of the negative wake have a more complicated behavior. 
It is illustrated by the left Fig.~3.6 where the spectra of different modes are shown together. 
It follows from them that the higher modes have a higher in absolute value thresholds at $\Delta Q=0$:
\begin{equation}
q_{\,0,-1}=-0.567\,Q_{s0},\;\;q_{-2,-3}=-3.46\,Q_{s0},\;\; q_{-4,-5}=-7.37\,Q_{s0}.
\end{equation}
All these thresholds increase in absolute value at increasing $\Delta Q$.
However, the lines of different modes move with different velocity, the mode $M_{0,-1}$ being faster of others.
Therefore it overtakes the mode $M_{-2,-3}$ at $\Delta Q = 6\,Q_{s0}$:
\begin{eqnarray}
q\,_{0,-1}=q_{-2,-3}=-9\,Q_{s0},\qquad q_{-3,-4}=-12\,Q_{s0}.
\end{eqnarray}
As a result, the mode $M_{-2,-3}$ stands the most unstable one at $\Delta Q\!>\!6\,Q_{s0}$. 
A further dependence of these thresholds on the SC tune shift is shown  in the right-hand Fig.~3.6.

It follows also from the figure that, in the TMCI threshold, tune of the mode $M_{-2,-3}$ satisfies the condition  $\,\nu_{th}\simeq-\Delta Q\,$ at $\,\Delta Q/Q_{s0}\gg1$.
Second Eq.~(3.22) obtains the form in the case:
\begin{equation}
q_{th} \,\simeq-\frac{{\cal Q}}{{\cal P}} \,\Delta Q
\end{equation}
However, the ratio  ${\cal Q}/{\cal P}$ changes when the marking point moves along the green line of Fig.~3.4, and the threshold is being achieved when the condition  $\,dq=0$ is fulfilled, that is $\,d{\cal P/P}=d\cal Q/Q$.
It is the point of tangency of the curve with the straight line 
 $\,{\cal P}=k{\cal Q}$ where $k$ is a constant.
This tangent is shown in Fig.~3.4 by the straight green line which shows  that $\,k=0.75$.
Therefore the asymptotic TMCI thresholds of the mode 
$M_{-2,-3}$ is: 
\begin{equation}
 q_{-2,-3}\rightarrow -1.33\,\Delta Q. 
\end{equation}
Using similar method, one can find threshold of the mode $M_{-4,-5}$ at large tune shift: $q_{-4,-5}\rightarrow -1.64\,\Delta Q$ (blue lines in Fig.~3.4).


\section{Synchrotron tune spread}


Eigenmodes of a bunch in the square well beyond the hollow bunch model are investigated in this section to estimate an effect of the synchrotron tune spread.  
Eq.~(3.2) with $q=0$ and boundary conditions (2.12) will be considered for this purpose.
It is easy to see that both the values $Y^{(+)}$ and $\bar Y$ are proportional to $\cos m\phi$ with an integer $m$ in the case.
Therefore the equation is converted to the following:
\begin{eqnarray}
Y^{(+)}(Q_s) = \frac{\hat\nu\Delta Q\,\bar Y}{\hat\nu^2-m^2 Q_s^2}
\end{eqnarray}
The synchrotron tune $Q_s\propto |u|$ is used here as the synchrotron phase conjugate variable. 
Then the relationship of the used functions can be taken in the form
\begin{eqnarray}
\bar Y = \int_0^\infty Y^{(+)}(Q_s)F(Q_s)\,dQ_s 
\end{eqnarray}
with the distribution function satisfying the normalization condition 
\begin{eqnarray}
\int_0^\infty F(Q_s)\,dQ_s=1
\end{eqnarray}
The substitution of these functions in Eq.~(3.28) and the integration results in the dispersion equation for the tune:  
\begin{eqnarray}
\hat\nu \Delta Q \int_0^\infty\frac{F(Q_s)\,dQ_s}
{\hat\nu^2-m^2Q_s^2}=1
\end{eqnarray}
Excluding the trivial case $m=0$, the equation can be represented in the form 
\begin{eqnarray}
\int_0^\infty\frac{F(Q_s)\,dQ_s}{Q_s-\hat\nu/m} + \int_{-\infty}^0\frac{F(-Q_s)\,dQ_s}{Q_s-\hat\nu/m} =-\frac{2m}{\Delta Q}
\end{eqnarray}
Let's introduce now the even normalized function $\Phi(s)$ where
$s=Q_s/Q_{s0}$ with $Q_{s0}$ as characteristic synchrotron frequency (e.g. maximal tune or rms spread).
It should be $\Phi(s)\propto F(sQ_{s0})$ at $s>0$ being normalized by the condition 
\begin{eqnarray}
\int_{-\infty}^\infty \Phi(s)\,ds=1
\end{eqnarray}
Then Eq.~(3.31) coincides with Eqs.~(1.20)-(1.21) up to the notation:
\begin{eqnarray}
\int_{-\infty}^\infty\frac{\Phi(s)\,ds}{z-s}=w_m,\qquad 
z = \frac{\hat\nu}{m Q_{s0}}, \qquad 
w_m =\frac{mQ_{s0}}{\Delta Q}
\end{eqnarray}
Therefore the results presented in Sec.~1.3 after Eq.~(1.23) are applicable here.
For example, for an uniform distributions of the tunes $0<Q_s<Q_{s0}$
\begin{equation}
z=\coth w_m \qquad {\rm that~is}\qquad 
\hat\nu_m=mQ_{s0}\coth \frac{mQ_{s0}}{\Delta Q}.
\end{equation}
Note that these eigentunes are real numbers exceeding the maximal synchrotron tune.


\chapter{Square bunch instability \\
(parabolic well, square bunch).}


Bunch of constant density (flat bunch, $\Delta Q=const$) in a parabolic potential well is considered in this chapter. 
Cartesian coordinates $\tau,\,u$ or the related cylindrical coordinates $A,\,\phi$ will be used:
\begin{equation} 
 \tau = A\cos\phi,\qquad u =-A\sin\phi,\qquad \phi=\Omega_0  Q_{s0} t
\end{equation}                         
Correspondent distribution function and linear density of the bunch are
\begin{equation} 
 F = \frac{1}{2\pi\sqrt{1-A^2}} = \frac{1}{2\pi\sqrt{1-\tau^2-u^2}}, \qquad\quad
 \rho(\tau)=\frac{1}{2} \;\; {\rm at} \;\; |\tau|<1
\end{equation}                         


\section{The bunch equation}


In the specified conditions, general equations of the bunch oscillations given by Eq.~(2.5) and Eq.~(2.9) have the form 
\begin{subequations}
\begin{equation}
\hat\nu Y + i\,Q_{s0}\frac{\partial Y}{\partial \phi} = \Phi(A\cos\phi)
\end{equation}
\begin{equation}
\Phi(\tau) = \Delta Q\bar Y(\tau) + q\int_0^{1-\tau}\bar Y(\tau+\tau') \exp(-i\zeta_\tau\tau')\,w(\tau')\,\,d\tau' 
\end{equation}
\begin{equation}
\bar Y(\tau) = \frac{1}{\pi} \int_{-\sqrt{1-\tau^2}}^{\sqrt{1-\tau^2}}
\frac{Y(\tau,u)\,du}{\sqrt{1-\tau^2-u^2}}  =  
\frac{1}{2\pi}\int_{0}^{2\pi}Y(\tau,\sqrt{1-\tau^2}\cos\eta)\,d\eta\quad
\end{equation}
\end{subequations}
where $\hat\nu\!=\!\nu\!+\!\Delta Q,\;\zeta_\tau\!=\!(\zeta\!-\!\nu)\theta_0$, 
and the last part of Eq.~(4.3c) is obtained by the replacement $u=\sqrt{1-\tau^2}\cos\eta$ in the previous part of this equation. 
A periodical solution of Eq.~(4.3a) is
\begin{eqnarray}
Y(A,\phi)=\frac{1}{Q_{s0}}\int_{0}^{2\pi}\Phi\big(A\cos(\phi+\psi)\big)\Psi(\hat\nu,\psi)\,d\psi
\end{eqnarray}
where
\begin{eqnarray}
\Psi(\hat\nu,\psi) = \frac{i\,\exp(-i\psi\hat\nu/Q_{s0})}
{1-\exp(-2\pi i\hat\nu/Q_{s0})} = \frac{\exp\big(i[\pi-\psi]\,\hat\nu/Q_{s0})\big)}
{2\sin\,\big(\pi\hat\nu/Q_{s0}\big)}
\end{eqnarray}
Using Eq.~(4.1), one can also represent the function $Y$ of Eq.~(4.4) in the form
\begin{eqnarray}
Y(\tau,u)=\frac{1}{Q_{s0}}\int_{0}^{2\pi}\Phi(\tau\cos\psi+u\sin\psi)\,\Psi(\hat\nu,\psi)\,d\psi.
\end{eqnarray}
Integration of this expression in $u$-domain, as it is described by the last part of Eq.~(4.3c), results in the relation of the functions $\bar Y(\tau)$ and $\Phi(\tau)$: 
\begin{eqnarray}
 \bar Y(\tau) = \frac{1}{2\pi Q_{s0}}
 \int_{0}^{2\pi}\Psi(\hat\nu,\psi)\,d\psi \int_0^{2\pi} \Phi(\tau\cos\psi+\sqrt{1\!-\!\tau^2}\sin\psi\cos\eta)\,d\eta. \quad
\end{eqnarray}
Substitution here the function $\Phi(\tau)$ from Eq.~(4.3b) provides the required integral equation for the function $\bar Y(\tau)$.
Its solution will be considered further in several stages.


\section{The bunch eigenmodes}


The first stage is determining the bunch eigenmodes without the wake,
that is only with the first term in the right-hand part of Eq.~(4.3b) 
being taken into account:
\begin{equation}
\Phi(\tau) = \Delta Q \,\bar Y(\tau)
\end{equation}
Then the integral equation (4.7) obtains the form
\begin{eqnarray}
 \bar Y(\tau) = \frac{\Delta Q}{2\pi Q_{s0}}\int_{0}^{2\pi}
 \Psi(\hat\nu,\psi)\,d\psi \int_0^{2\pi} 
 \bar Y(\tau\cos\psi+\sqrt{1\!-\!\tau^2}\sin\psi\cos\eta)\,d\eta. \quad
\end{eqnarray}
We will use an expansion of the functions $\bar Y$ in series of the (unnormalized) Legendre polynomials, writing it as
\begin{eqnarray}
\bar Y(\tau) = \sum_{n=0}^\infty y_n P_n(\tau)
\end{eqnarray}
As a result, the integral equation (4.9) for the function $\bar Y$ is replaced by an infinite series of equations for the coefficients $y_n$:
\begin{eqnarray}
y_k=\sum_{n=0}^\infty S_{k,n} y_n
\end{eqnarray}
with the matrix
\begin{eqnarray}
S_{k,n} = \frac{(2k+1)\Delta Q}{4\pi Q_{s0}} 
\int_{0}^{2\pi} \!\!\!\Psi(\hat\nu,\psi)\,d\psi \hspace{22mm}  \nonumber \\
\times \,\int_{-1}^1 \! P_k(\tau)\,d\tau \int_{-\pi}^\pi \! P_n(\tau\cos\psi+\sqrt{1\!-\!\tau^2}\sin\psi\cos\eta)\,d\eta. \quad
\end{eqnarray}
Next we will apply the Legendre addition theorem including the associated Legendre polynomials $P_n^m$ \footnote{E. Janke, F. Emde, and F. Losch, "Tafeln Hoherer Funktionen" XII/6.2, B. G. Verlagsgesellschaft, Stuttgart, 1960.} :
\begin{eqnarray}
P_n\big(\cos\alpha\cos\psi+\sin\alpha\sin\psi\cos\eta)
= P_n(\cos\alpha)P_n(\cos\psi)   \nonumber   \\          
+\;2\sum_{m=1}^n\frac{(n-m)!}{(n+m)!}P_n^m(\cos\alpha)P_n^m(\cos\psi)\cos m\eta
\quad {\rm at}\quad n \ge 1.
\end{eqnarray}
Substituting this relation with $\cos\alpha=\tau$ in Eq.~(4.12), and taking into account orthogonality of the Legendre polynomials, one can show that $\bf S$ is a diagonal matrix with 
\begin{eqnarray}
 S_{n,n}  = \frac{\Delta Q}{Q_{s0}} \int_{0}^{2\pi}\Psi(\hat\nu,\psi) P_{n}(\cos\psi)\,d\psi 
 \end{eqnarray}
It means that that the Legendre polynomials $\,P_n(\tau)\,$ are the eigenfunctions of Eq.~(4.9), and corresponding eigentunes satisfy the equation 
$\;S_{n,n}(\hat\nu)=1$.
Using Eq.~(4.5) and Eq.~(4,14), one can represent this equation in the form
\footnote{ F. Sacherer, Report No. CERN-SI-BR-72-5, 1972. }
\begin{eqnarray}
 1 = \frac{i\,\Delta Q }{Q_s[1-\exp(-2\pi i\hat\nu/Q_{s0})]}    \int_{0}^{2\pi} P_n(\cos\psi) \exp(-i\psi\hat\nu/Q_{s0}) \,d\psi 
 \end{eqnarray}
Applying further the trigonometric representation of the Legendre polynomials
\footnote{E. Janke, F. Emde, and F. Losch, "Tafeln Hoherer Funktionen" XII/2.1,
B. G. Verlagsgesellschaft, Stuttgart, 1960.}
\begin{eqnarray}
P_n(\cos\psi)=\sum_{m=0}^{n}C_{n,m}\cos m\psi 
\end{eqnarray}
one can bring Eq.~(4.15) into the form 
\begin{eqnarray}
 1 = \Delta Q\sum_{m=0}^{n}\frac{C_{n,m}\hat\nu}{\hat\nu^2-m^2 Q_{s0}^2}.
 \end{eqnarray}
Actually the coefficients $C_{n,m}$ are nonzero if the indexes $n$ and $m$ 
are of the same parity.
Therefore Eq.~(4.17) is an algebraic equation of power $n+1$.
It has $n+1$ roots which will be denoted farther as $\hat\nu_{nm}$ with $\,m=-n,\,-n+2,\dots ,\,n-2,\,n$.
Explicit forms of these algebraic equations are
\begin{subequations}
\vspace{1mm}
\begin{eqnarray} 
\hat\nu=\Delta Q \qquad{\rm at}\qquad n=0,  
\end{eqnarray}
\vspace{-6mm}
\begin{eqnarray}
\hspace{-3mm}\hat\nu \prod\nolimits_{k=1}^{n/2} [\hat\nu^2\!-\!(2k)^2Q_{s0}^2] = \Delta Q \prod\nolimits_{k=1}^{n/2} [\hat\nu^2\!-\!(2k\!-\!\!1)^2Q_{s0}^2] 
\qquad\;\;\;\; {\rm at~even~}n,\quad
\end{eqnarray}
\vspace{-5mm}
\begin{eqnarray}
\!\!\hat\nu \prod\nolimits_{k=1}^{(n+1)/2} [\hat\nu^2\!-\!(2k\!-\!\!1)^2Q_{s0}^2] = \Delta Q \prod\nolimits_{k=0}^{(n-1)/2} [\hat\nu^2\!-\!(2k)^2Q_{s0}^2]
\;\, {\rm at~odd~} n. \quad\;
\end{eqnarray}
\end{subequations}
Their solutions are plotted against the SC tune shift in the left-hand panel of Fig.~4.1.
\begin{figure}[t!]
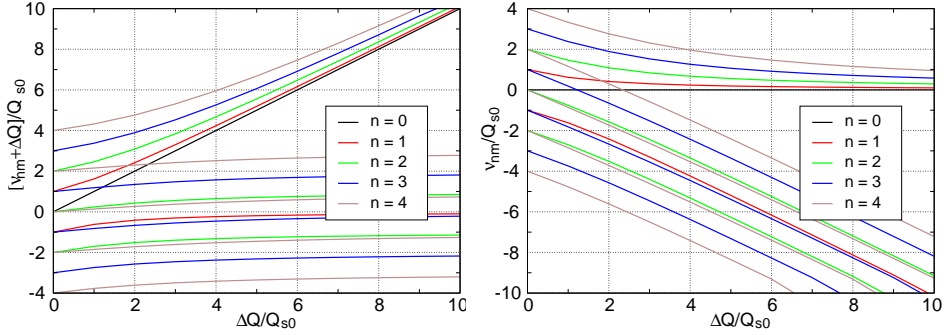

\hspace{2mm}
\begin{minipage}[h!]{0.45\linewidth}
\centerline{\epsfig{file=4_1a_tune.eps,width=\linewidth}}
\end{minipage}
\begin{minipage}[h!]{0.45\linewidth}
\centerline{\epsfig{file=4_1b_tune.eps,width=\linewidth}}
\end{minipage}
\caption{Flat bunch in the parabolic potential well. Different views of the bunch eigentunes are given being plotted against the SC tune shift referred to the synchrotron tune $Q_{s0}$. 
Several lowest multipoles are shown.}
\end{figure}
At $\Delta Q=0$, the tunes have the form $\nu_{n,m}=\hat\nu_{n,m}=mQ_{s0}$ independently on $n$, what means that the eigenfunctions are the multipoles $Y\!\propto\! \exp(im\phi)$ in the case.
A lot of tunes with different $n$ take start from each point $mQ_{s0}$.
At $\Delta Q\ne 0$, related eigenfunctions $Y_{n,m}$ appear as a combination of the multipoles with different coefficients, dependent on the amplitude $A$.

It is seen that all the eigentunes form two different clusters which are distinctly separated in both panels of Fig.~4.1.
Only the tunes with the highest multipole indexes $m=n$ at each $n$ form the higher cluster being rather well described by the formula 
\begin{equation}
\hat\nu_{n,n} \simeq \frac{\Delta Q}{2} 
+\sqrt{\frac{\Delta Q^2}{4}+\frac{n(n+1)Q_{s0}^2}{2}}
\rightarrow \Delta Q+\frac{n(n+1)Q_{s0}^2}{\Delta Q}
\end{equation}
where the ultimate value at $\Delta Q/Q_{s0}\rightarrow\infty$ is shown as well.

In contrast with that, other modes like $m<n$ are located in the lower cluster. 
These tunes only slightly depend on $n$ according to the approximate expression:  
\begin{equation}
\hat\nu_{n,m} \simeq (m+1) Q_{s0}-\frac{Q_{s0}^2}{\Delta Q}
\qquad {\rm{at}} \qquad -n\le m \le n-2
\end{equation}
The corresponding values $\nu_{n,m}=\hat\nu_{n,m}-\Delta Q$ are plotted as well in the right-hand panel of Fig.~4.1. It allows to make a detailed comparison with the square well model which has been shown in Fig.~3.1.
It is evident the very close resemblance of the figures at $|m|=n$: $\nu_{n,n}\simeq \nu_{+m}$, and $\nu_{n,-n}\simeq \nu_{-m}$. 
Note that the relations are exact  at $n=0$ and $n=1$.
The modes $|m|\ne n$ are absent at all in the case of the rectangular well.

When the function $\bar Y_n(\tau)$ is known, the eigenfunctions $Y_{n,m}$ can be found with help of Eq.~(4.4) or Eq.~(4.6)  where the eigenfunctions $\bar Y(\tau)=P_n(\tau)$ and the eigentunes $\hat\nu=\hat\nu_{n,m}$ have to be substituted resulting in
\begin{eqnarray}
 Y_{n,m}(A,\phi) = \frac{\Delta Q}{Q_{s0}}\int_{0}^{2\pi} P_n\big(A\cos(\phi+\psi)\big)\Psi(\hat\nu_{n,m},\psi)\,d\psi 
\end{eqnarray}
The function $\Psi$ is generally given by Eq.~(4.5); however, it is more convenient here to use the notation $\hat\nu_{n,m}=(K+x)Q_{s0}$, with an integer $K$ and $0<x<1$, the rewrite the relations in the form
\begin{eqnarray}
 Y_{n,m} = \int_{0}^{2\pi K} H(\xi)\exp(-i\xi)\,d\xi
 \hspace{30mm}
\end{eqnarray}
where
\begin{eqnarray}
 H(\xi) = \frac{i\,\Delta Q\,P_n\big(A\cos(\phi+\xi/K)\big)\,\exp(-ix\xi/K)}
{Q_{s0} N\,[1-\exp(-2\pi i x)]}
\end{eqnarray}
Simple transformations of Eq.~(4.22a) result in
\begin{eqnarray}
 Y_{n,m} = \sum_{k=0}^{K-1}\int_{2\pi k}^{2\pi(k+1)} \hspace{-5mm}
 H({\xi})\exp(-i\xi)\,d\xi 
 = \sum_{k=0}^{K-1}\int_0^{2\pi} \hspace{-3mm}H(2\pi k+\xi)
 \exp(-i\xi)\,d\xi\quad
\end{eqnarray}
Let's apply this expression to find the function $Y_{n,n}$ at $\Delta Q/Q_{s0}\gg 1$.
As it follows from Fig.~4.1 and Eq.~(4.19),  $\hat\nu_{n,n}\simeq\Delta Q\gg Q_{s0}$, that is $K\gg 1$ in the case.
Then, according to Eq.~(4.22b), $H(\xi)$ is a slowly changed function, and its decomposition to the Taylor series allows to get from Eq.~(4.23) the approximate expression:
\begin{eqnarray}
 Y_{n,n} \simeq 2\pi i\sum_{k=0}^{K-1} H'(2\pi k)
 \simeq i\sum_{k=0}^{K-1}(H_{k+1}-H_k)=i\,(H_K-H_0)
\end{eqnarray}
where $H_k=H(2\pi k)$.
Taking into account Eq.~(4.23) and using the relations 
$NQ_s\simeq \hat\nu \simeq\Delta Q$, obtain finally 
\begin{eqnarray}
 Y_{n,n}(A,\phi)\simeq \frac{\Delta Q}{Q_{s0} N} P_n(A\cos\phi)\simeq P_n(A\cos\phi) = \bar Y_n(\tau)
\end{eqnarray}


\section{Solution by an expansion}


Now we consider Eq.~(4.7) with the function $\Phi(\tau)$ including a wake field 
according Eq.~(4.3b).
We will find its solution using the expansion of the functions in series on the Legendre polynomials, like Eq.~(4.10):
\begin{eqnarray}
\bar Y(\tau) = \sum_{n=0}^\infty y_n P_n(\tau),\qquad
\Phi (\tau) = \sum_{n=0}^\infty f_n P_n(\tau)
\end{eqnarray}
A substitution of these expressions to Eq.~(4.7) and further transformations like to those made in the previous section (with the replacement $\,\Delta Q y_n\Rightarrow f_n$ if needed) provides the relation of the coefficients:
\begin{eqnarray}
f_n=Q_s {\cal F}_n(\hat\nu)\, y_n
\end{eqnarray}
where
\begin{eqnarray}
{\cal F}_n(\hat\nu) = \left[\int_{0}^{2\pi}
\Psi(\hat\nu,\psi) P_n(\cos\psi)\,d\psi\right]^{-1}
\end{eqnarray}
With the function $\Psi$ given from Eq.~(4.5),
the last expression is the real function
\begin{eqnarray}
{\cal F}_n(\hat\nu) = (-1)^n\sin\frac{\pi\hat\nu}{Q_{s0}}
\left[\int_0^\pi\cos\left(\frac{\hat\nu\psi}{Q_{s0}}\right)
P_n(\cos\psi)\,d\psi\right]^{-1}
\end{eqnarray}
One more relation of the coefficients can be obtained by a substitution of Eq.~(4.27) to Eq.~(4.3b):
\begin{eqnarray}
 f_n = \Delta Q\,y_n \hspace{44mm} \\ \nonumber  + \;
 q\,\frac{2n+1}{2} \sum_{n'=0}^\infty  y_{n'} \int_{-1}^{1}
 P_n(\tau)\,d\tau \int_0^{1-\tau}\!\!\!P_{n'}(\tau+\tau') \,
 \exp(-i\zeta_\tau\tau')\,d\tau'
\end{eqnarray}
The interplay of Eqs.~(4.28) and (4.31) provides the series of equations for the coefficients $y_n$:
\begin{equation}
 \left( {\cal F}_n (\hat\nu) - \frac{\Delta Q}{Q_{s0}}\right)y_n 
= \frac{q}{Q_{s0}} \sum_{n'=0}^\infty R_{n,n'}y_{n'}
\end{equation}
with the matrix
\begin{eqnarray}
R_{n,n'} = \frac{2n+1}{2} \int_{-1}^{1} P_n(\tau)\exp(i\zeta_\tau\tau)\,d\tau \int_\tau^{1}\!\!\!P_{n'}(\tau)\,\exp(-i\zeta_{\tau}\tau')\,d\tau'
\end{eqnarray}


\subsection{Head-tail instability}


An interaction of different modes is negligible at $|q|\ll Q_{s0}$, so the series of equations (4.32) decomposes in the separate equations
\begin{eqnarray}
{\cal F}_n(\hat\nu) = \frac{\Delta Q + q R_{n,n}}{Q_{s0}}
\end{eqnarray}
\begin{figure}[b!]
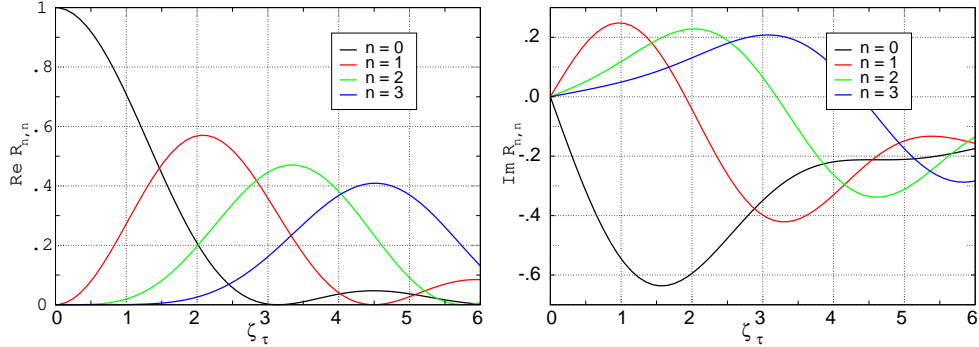

\hspace{2mm}
\begin{minipage}[h!]{0.47\linewidth}
\centerline{\epsfig{file=4_2_real.eps,width=\linewidth}}
\end{minipage}
\begin{minipage}[t!]{0.47\linewidth}
\centerline{\epsfig{file=4_2_imag.eps,width=\linewidth}}
\end{minipage}
\caption{Normalized tunes of different head-tail modes of a bunch in the parabolic potential well with a flat wake, vs chromaticity. 
The real/imaginary parts are shown in the left/right panels being the even/odd functions of chromaticity.}
\end{figure}
At $q=0$, this equation turns into Eq.~(4.14) for the bunch eigentunes.
As it has been shown in Sec.~4.2, the equation has $n+1$ solutions some of which are plotted in Fig.~4.1 being marked as $\hat\nu_{n,m}$.
Therefore, at $|qR_{n,n}|\ll\Delta Q$, additions to the tunes are:
\begin{eqnarray}
\delta\hat\nu_{n,m}  \simeq \frac{q R_{n,n}}{Q_{s0}}  \frac{d\hat\nu_{n,m}} {d{\cal F}} = q R_{n,n} \frac{d\hat\nu_{n,m}} {d(\Delta Q)} 
\end{eqnarray}
The last derivative can be found as a slope of the curve in the left-hand panel of Fig.~4.1. 
It is seen that the slopes are small for all modes like $\hat\nu_{n,m\ne n}$; however, 
they are rather large for the modes  $\hat\nu_{n,n}$ which occur 
the most sensitive to the wake field. 
Corresponding additions to the tune follows from Eq.~(4.19): 
\begin{equation}
\delta\hat\nu_{n,n} = \frac{qR_{n,n}}{n\!+\!1} \left(\frac{n\Delta Q} {\sqrt{\Delta Q^2+(n\!+\!1)^2Q_{s0}^2}}+1\right) \simeq qR_{n,n}
\end{equation}
Real and imaginary parts of the matrix elements $R_{n,n}$ are plotted in Fig.~4.2 against the chromaticity factor at different $n$. 
The similarity of Figs.~3.2 and 4.2 becomes obvious, if the relation 
$\zeta_\phi=2\zeta_\tau/\pi$ is taken into account. 


\subsection{TMCI: multimode solution. }


Solutions of Eq.~(4.32) without chromaticity is considered in this section.
It is assumed that the series of equations is truncated by the assumption $y_n=0$ an $n>N$. 
All details have been set out in Sec.~3.4, and the only difference is a specific form of the matrix {\bf T} which is in the case.
\begin{equation}
T_{n,n'} = q R_{n,n'}+(\Delta Q-Q_s{\cal F}_n)\,\delta_{n,n'}
\qquad{\rm at}\qquad 0\le n, n' \le N.
\end{equation}
The matrix {\bf R} is generally given by Eq.~(4.33) but it looks much simpler at $\zeta_\tau=0$.
Using the relation
\begin{equation}
(2n+1)P_n(\tau) = \frac{d\,[P_{n+1}(\tau)-P_{n-1}(\tau)]}{d\tau}
\end{equation}
and the normalization conditions 
\footnote{E. Janke, F. Emde, and F. Losch, "Tafeln Hoherer Funktionen" XII/6.5, B. G. Verlagsgesellschaft, Stuttgart, 1960.}
obtain
\begin{equation}
R_{n,n'} = \frac{ \delta_{n',n+1} - \delta_{n',n-1}} {2n'+1}.
\end{equation}
A small fragment of the matrix is represented in Table 4.1.

Results of the calculations are very similar to those presented in the previous chapter for the square potential well.
At $\,q>0$, the TMCI threshold almost does not depend on the number $N$ appearing as a monotone decreasing function of $\Delta Q$ like Eq.~(3.17)
\begin{eqnarray}
q_{th}\simeq \frac{0.57 Q_{s0}}{\sqrt{1+(1.2\,\Delta Q/Q_{s0})^2}}
\end{eqnarray}

\begin{table}[t!]
\begin{center}
\caption{Fragment of the matrix $\bf R$}
\vspace{2mm}
\begin{tabular}{|c|c|c|c|c|c|c|}
\hline 
$n'\rightarrow$  &~0~~~&~~1~~&~~2~~&~~3~~&~~4~~&~~5     \\
\hline
$  ~ n=0 ~    $  &  1  & 1/3 &  0  &  0  &  0  &  0     \\
$  ~ n=1 ~    $  & -1  &  0  & 1/5 &  0  &  0  &  0     \\   
$  ~ n=2 ~    $  &  0  &-1/3 &  0  & 1/7 &  0  &  0     \\   
$  ~ n=3 ~    $  &  0  &  0  &-1/5 &  0  & 1/9 &  0     \\   
$  ~ n=4 ~    $  &  0  &  0  &  0  &-1/7 &  0  &  1/11   \\  
$  ~ n=5 ~    $  &  0  &  0  &  0  &  0  &-1/9 &  0      \\   
\hline
\end{tabular}
\end{center}
\end{table}
\begin{figure}[h!]
\vspace{3mm}
\centerline{\epsfig{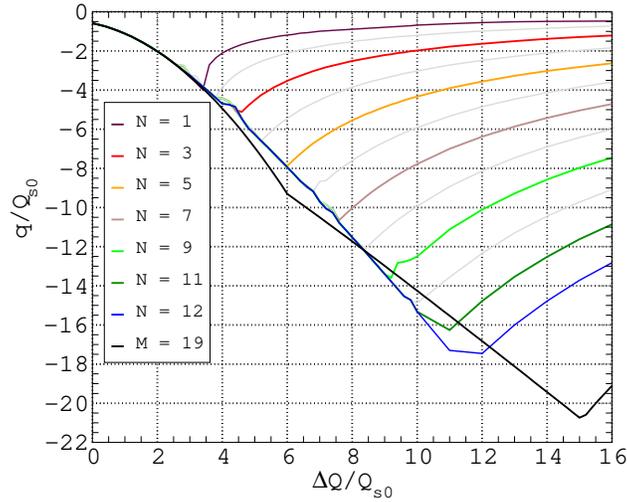}}
\caption{The TMCI threshold of a bunch of constant density with a flat wake in a parabolic potential well (linear synchrotron oscillations), against the SC tune shift.
Different curves are obtained by solution of the equation $\det\bf T=0$ where the matrix $\bf T$ is given by Eq.~(4.39) with different $\,N$.
The drop-down part of each curve is the region of its applicability, which expands when  $\,N$ is increasing.
The rising lines do not confirm each other and should be ignored.
The black line is copied from Fig.~3.3 for a comparison.}
\vspace{11mm}
\end{figure}
The threshold of negative wake has a more complicated behavior 
which is shown in Fig.~4.3, looking like Fig.~3.3. 
In both cases, there is the drop-down part which belongs to the approximations with different $N$ ($M$ in Chapter 3).
Any approximation with the higher $n$ follows the course which the lower ones have charted, providing the continuation to the higher $\,\Delta Q$. 
The coming back lines do not confirm each other and cannot be treated as credible results. 
Actually, the break of any line means the end of the approximation applicability with given $N$.
The black line in the figure is taken from Fig.~3.3 for a comparison. 
It is seen that the thresholds are rather close in the square well and in the parabolic one, if the bunch length is the same in the cases.


\section{TMCI: numerical solution.}


According to Sec.~2, the bunch eigenfunctions $\bar Y_n$ are the Legendre polynomials $P_n(\tau)$ that is they must satisfy the equation:    
\begin{eqnarray}
\frac{d}{d\tau}\left(\frac{1\!-\!\tau^2}{2} \frac{d\bar Y_n}{d\tau}
\right) + \frac{n(n+1)}{2} \bar Y_n = 0.
\end{eqnarray}
However, it provides no information about the bunch eigentunes, so the dispersion equation (4.15) should be added to complete the question.
At each $\,n\,$, the equation has $\,n+1\,$ solutions with different $\,m\,$ which approximate forms are given by Eqs.~(4.19)-(4.20) and Fig.~(4.1).
According them, the highest tunes $\,\hat\nu_{n,n}$ stand apart from others 
being given by Eq.~(4.19) at $\Delta Q/Q_{s0}\gg1$.
Therefore, one cane take for these tunes:
\begin{eqnarray}
\frac{\nu\hat\nu}{Q_{s0}^2} = \frac{n(n+1)}{2}.
\end{eqnarray}
Correspondingly, Eq.~(4.41) obtains the form: 
\begin{eqnarray}
\frac{d}{d\tau}\left(\frac{1\!-\!\tau^2}{2} \frac{d\bar Y_n}{d\tau}
\right) + \frac{\nu\hat\nu}{Q_{s0}^2} \bar Y_n = 0.
\end{eqnarray}

It is needless to say that this equation can be considered 
only as a model requiring an additional validation. 
We continue the investigation including the flat wake and representing the result like Eq.~(3.19) related the square well model.
At zero chromaticity, the equation is
\begin{eqnarray}
\frac{1-\tau^2}{2}\,\bar Y^{''}(\tau)=\tau \bar Y'(\tau)-{\cal P} \bar Y(\tau) + {\cal Q}\int_\tau^1 \bar Y(\tau_1)\,d\tau_1
\end{eqnarray}
where the sign $'$ means the derivative with $\tau$, and the notations like Eq.~(3.20) are used:
\begin{eqnarray}
{\cal P}=\frac{\nu\hat\nu}{Q_{s0}^2}=\frac{\hat\nu(\hat\nu-\Delta Q)}{Q_{s0}^2},  \qquad {\cal Q} = \frac{q\,\hat\nu}{Q_{s0}^2}
\end{eqnarray}
The boundary conditions for the equation follow from itself because its right-hand part should be zero at $\tau=\pm 1$:
\begin{eqnarray}
\bar Y'(1) = {\cal P}\bar Y(1), \qquad \bar Y'(-1) =-{\cal P} \bar Y(-1) 
+ {\cal Q}\int_{-1}^1 \bar Y(\tau)\,d\tau
\end{eqnarray}
 \begin{figure}[t!]
 \centerline{\epsfig{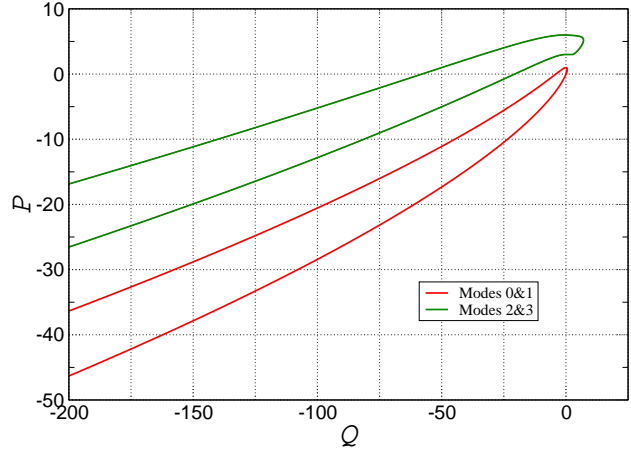}}
 \vspace{3mm}
 \caption{Real eigennumbers of Eq.~(4.44) with boundary conditions Eq.~(4.42). Two lowest loops are shown.}
\end{figure}

The equation can be solved numerically like it has been done in Sec.~3.5 for the square potential well. 
At any real ${\cal Q}$, there is a lot of eigennumbers ${\cal P}_n$ which satisfy  the  boundary conditions. 
Some of them are plotted in Fig.~4.4 which looks very similar to Fig.~3.4 related to the square well. 
As expected, at $\cal Q\!=\,$0, the values ${\cal P}_n$ are $n(n+1)/2$ with $n=0,\,1,\,2,\,3$, and the loops appear due to a coalescence of branches.
The bunch tunes calculated by using of Eq.~(4.45) are 
\begin{equation}
\hat\nu_{n\pm} = \frac{\Delta Q}{2} \pm \sqrt{\frac{\Delta Q^2}{4} 
+{\cal P}_n Q_{s0}^2}  \qquad {\rm at} \qquad
q = \frac{Q_{s0}^2{\cal Q}}{\hat\nu_{n\pm}}  
\end{equation}

The graphs of these functions look similar to those shown in the left and center panels of Fig.~3.5.
The turning point of any curve shows the instability threshold of corresponding mode for a given $\Delta Q$.
If $q>0$, the points are located in the region $\,\hat\nu>0$, that is the instability arises due to the coalescence of the tuns $\hat\nu_{n+}$ described by the Eq.~(4.47) with different $\,n$.
The threshold of the most unstable mode is described well by Eq.~(4.40) obtained by the expansion method:
$$
q_{th}\simeq \frac{0.57 Q_{s0}}{\sqrt{1+(1.2\,\Delta Q/Q_{s0})^2}}
$$
\begin{figure}[t!]
\centerline{\epsfig{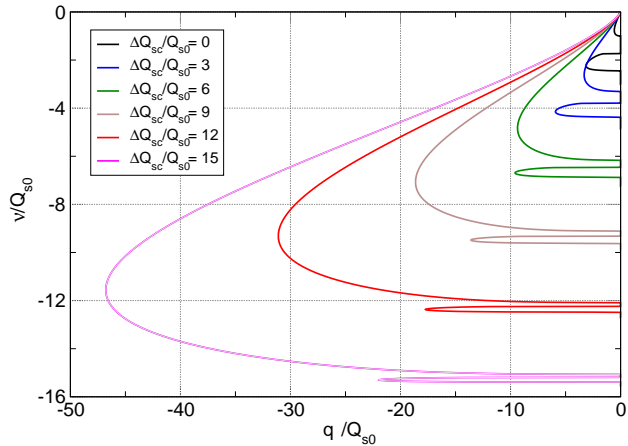}}
\caption{The tunes of the bunch against the wake strength at different space charge tune shifts obtained with help of Fig.~4.4 and Eq.~(4.46).
The tunes of the modes $m=(0-1)$, and $m=(2-3)$ are plotted at $q<0$. The higher mode has a higher absolute value tune shift.
The turning point of each loop indicates the TMCI threshold of the corresponding coupling mode.}
\end{figure}

With the negative wake, the bunch tunes  are plotted in Fig.~4.5 which  looks very like to the left-hand panel of Fig.~3.6.
Thresholds of tho modes are plotted in the case in Fig.~4.6 against the SC tune shift by the red and green solid lines.
The dashed lines in the figure are the TMCI threshold of the same modes of the bunch in the square potential well (the red and green lines in Fig.~3.6).
The blue line is taken from Fig.~4.3 where it has been obtained with help of the expansion technique at $N=12$. 
All these results are in a good agreements.

\begin{figure}[t!]
\vspace{11mm}
\centerline{\epsfig{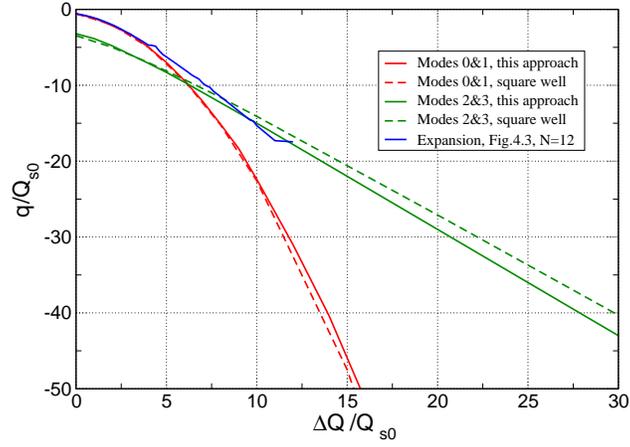}}
\caption{The TMCI thresholds of a bunch of constant density in parabolic potential well with the negative wake.
The thresholds of different modes against the space charge tune shift. 
The red and blue lines refer to the TMCI modes modes $M_{0,-1}$ and $M_{-2,-3}$.
The dashed lines are placed for a comparison being obtained in Chapter 3 in the framework of the square well model with a hollow bunch.}
\vspace{5mm}
\end{figure}


\chapter{Arbitrary bunch in the parabolic well, flat wake} 

Investigation of a bunch in the parabolic potential well [see Eq.~(4.1)] will be continued in this chapter for general bunch forms besides the flat one.

\section{The bunch integral equation}


Equations (4.3a-b), (4.4), and (4.6) are acceptable with any bunch shape; however, Eq.~(4.5) should be upgraded to take into account a possible dependence of the SC tune shift on the longitudinal coordinate.
As a result, the full set of equations obtains the form
\begin{subequations}
\begin{eqnarray}
Y(\tau,u)=\frac{1}{Q_{s0}}\int_{0}^{2\pi}\Phi(\tau\cos\psi+u\sin\psi)\,
\Psi(\tau,u,\psi)\,d\psi,\;\;\;
\end{eqnarray}
\vspace{-5mm}
\begin{eqnarray}
\Psi(\tau,u,\psi) = \frac{i\,\exp\left(-\frac{i}{Q_{s0}}\int_0^\psi
\hat\nu\big(\tau\cos\gamma+u\sin\gamma\big)d\gamma\right)} 
{1-\exp\left(-\frac{i}{Q_{s0}} \int_0^{2\pi}
\hat\nu\big(\sqrt{\tau^2+u^2}\cos\gamma\big) d\gamma \right)},
\end{eqnarray}
\vspace{-4mm}
\begin{eqnarray}
\hat\nu(\tau)=\nu+\Delta Q(\tau).  \hspace{62mm}
\end{eqnarray}
\end{subequations}
where $\nu$ is the coherent tune, and $\Delta Q(\tau)$ is the SC tune shift dependent on the longitudinal coordinate.
The relation of the functions $Y(\tau,u)$ and $\bar Y(\tau)$ is determined by Eq.~(2.9) which includes the bunch distribution function $F$.
For a clarity and an comparison of obtained results with with the previous ones, we will use as an example the function  
\begin{subequations}
\begin{eqnarray}
F=\frac{2\alpha+1}{2\pi}(1-A^2)^{\alpha-1/2} = \frac{2\alpha+1}{2\pi} (1-\tau^2-u^2)^{\alpha-1/2},  
\end{eqnarray}
\vspace{-7mm}
\begin{eqnarray}
\lambda(\tau) =\int F\,du = C_\alpha(1-\tau^2)^\alpha.
\end{eqnarray}
\end{subequations}
with an arbitrary power $\alpha$ and the normalized coefficients $C_\alpha$ some of which are given in Table~5.1 
(note that $\alpha=0$ in Ch.~4).
\begin{table}[b!]
\begin{center}
\vspace{-9mm}
\caption{Normalizing coefficients}
\vspace{3mm}
\begin{tabular}{|c|c|c|c|c|c|}
\hline
$\alpha  $  &~0~~~&~1/2~&~~1~~&~3/2~~&~~2~~ \\
\hline
$C_\alpha$  & 1/2 & $2/\pi$ & 3/4 & $8/3\pi$ & 15/16\\
\hline
\end{tabular}
\end{center}
\end{table}
Then the mentioned relation is
\begin{eqnarray}
\bar Y(\tau)=\frac{\int_0^{\pi}Y(\tau,\sqrt{1-\tau^2}\cos\eta) \sin^{2\alpha}\!\eta\,d\eta} {\int_0^{\pi}\sin^{2\alpha}\!\eta\,d\eta}
\end{eqnarray}
Substituting here Eq.~(5.1a) obtain
\begin{eqnarray}
\bar Y(\tau)=\frac{1}{Q_{s0}\int_0^\pi \sin^{2\alpha}t\,dt}\hspace{44mm}\\\times
\int_0^{\pi}\!\!\!\sin^{2\alpha}\eta\,d\eta 
\int_{0}^{2\pi}\!\!\!\Phi(\tau\cos\psi\!+\!\sqrt{1\!-\!\tau^2}\sin\psi
\cos\eta)\, \Psi(\tau,\sqrt{1\!-\!\tau^2}\cos\eta,\psi)\,d\psi\nonumber
\end{eqnarray}
A general view of the function $\Phi(\tau)$ is given by Eq.~4.3b.
In particular, $\Phi(\tau)=\Delta Q(\tau)\bar Y(\tau)$ if the wake-field absences, providing the closed equation for the bunch eigenmodes. 
A substitution of this expression in Eq.~(5.4), at the distributions given by Eq.~(5.2), results in the equation:
\begin{eqnarray}
\Phi(\tau)=\frac{\Delta Q_0\,(1-\tau^2)^\alpha}{Q_{s0}\int_0^\pi \sin^{2\alpha}t\,dt}\hspace{44mm}\\\times
\int_0^{\pi}\!\!\!\sin^{2\alpha}\eta\,d\eta 
\int_{0}^{2\pi}\!\!\!\Phi(\tau\cos\psi\!+\!\sqrt{1\!-\!\tau^2}\sin\psi
\cos\eta)\, \Psi(\tau,\sqrt{1\!-\!\tau^2}\cos\eta,\psi)\,d\psi\nonumber
\end{eqnarray}
where $\Delta Q_0$ is the space charge tune shift in the bunch center.
The function $\Psi(\tau,u,\psi)$ is determined by Eq.~(5.1b) with $u=\sqrt{1-\tau^2}\cos\eta$.


\section{Are the coherent oscillations damping at a large space charge?}


It is pertinent to recall that the above equations are obtained using the Laplace transform at Im $\,\nu>0$.
The denominator of Eq.~(5.1b) is a nonzero number in the case at any synchrotron amplitude $A=\sqrt{\tau^2+u^2}$.
However, a simple use of the equation with a real $\nu$ is possible only if the value $\nu+\overline{\Delta Q(A)}$ is a non-integer number at any $A$ with $\overline{\Delta Q(A)}$ as the average SC tune shift a the particle with the synchrotron amplitude $A$: 
\begin{equation}
\overline{\Delta Q(A)} = \frac{1}{2\pi}\int_0^{2\pi}\Delta Q(A\cos\gamma)\,d\gamma.
\end{equation}
For example, $\Delta Q_0/2<\Delta Q(A)<\Delta Q_0$ at $\alpha=1$, that is the above condition is executable at least some $\,\nu\,$ only at $\Delta Q_0<2$.
The similar statement is valid at any $\alpha\ne0$.

It would be possible to conclude from this that a simple use of Eq.~(5.1b) is unacceptable at Im $\,\nu\le0$, and a bypass of an appearing pole is required for an analytic continuation of the expressions like Eq.~(1.29).
Then all eigenmodes of the bunch would be damping ones at sufficiently large 
$\Delta Q$ (Landau damping. see Sec.~1.3).

However, the conclusion requires an examination because the nominator of Eq.~(5.5) can also be zero at these conditions.  
Because the equation fails to solve at $\alpha\ne0$, results of Ch.~4 will be engaged for the discussion.

The main of them is that all the eigentunes are distributed in two clusters separated by the interval $\sim\Delta Q$. 
The eigentunes $\nu_{n,n}\rightarrow 0$ at $\Delta Q\rightarrow\infty$ in the upper cluster, and $\nu_{n,m}\simeq-m\Delta Q$ in the lower one (Fig.~4.1).
The assumption is that the same statements are valid for any distributions, for example, for given by Eq.~(5.2) with an arbitrary $\alpha$. 
\\

To explore the upper cluster, we we return to Eq.~(5.1) rewriting them in terms of $(A,\phi)$: 
\begin{subequations}
\begin{eqnarray}
Y(A,\phi)=\frac{1}{Q_{s0}}\int_{0}^{2\pi}\bar Y\big(A\cos(\phi+\psi)\big)\,
\Delta Q\big(A\cos(\phi+\psi)\big)\,
\Psi(A,\phi,\psi)\,d\psi,\quad
\end{eqnarray}
\vspace{-5mm}
\begin{eqnarray}
\Psi(A,\phi,\psi) = \frac{i\,\exp\left(-\frac{i}{Q_{s0}}\int_0^\psi\Delta Q\big(A\cos(\phi+\gamma)\big)\,d\gamma-i\nu\psi/Q_{s0} \right)} 
{1-\exp\left(-\frac{i}{Q_{s0}}\int_0^{2\pi}
\Delta Q\big(A\cos\gamma\big) d\gamma-2\pi i\nu/Q_{0}s\right)} \hspace{17mm}
\end{eqnarray}
\end{subequations}
The last equation includes the "suspicious" denominator, and  the question is, 
can the numerator compensate for its zeros.
It is convenient to use the new variable $\xi$ instead of $\psi$ at $\Delta Q/Q_{s0}\gg1$: 
\begin{eqnarray}
\xi =\frac{K}{(K+x)Q_{s0}}\int_0^\psi\Delta Q \big(A\cos(\phi+\gamma)\big)\,d\gamma
\end{eqnarray}
where the positive integer number $K\gg1$, and the parameter $0<x<1$ is added 
to ensure the relation 
\begin{eqnarray}
\xi(2\pi)=\frac{K}{(K+x) Q_{s0}}\int_0^{2\pi}\Delta Q(A\cos\gamma)\,d\gamma=2\pi K
\end{eqnarray}
Then Eq.~(5.7) is reducible to the form like Eq.~(4.22)
\begin{eqnarray}
Y(A,\phi) = \int_{0}^{2\pi K} H(A,\phi,\xi)\exp(-i\xi)\,d\xi\quad
\end{eqnarray}
where
\begin{eqnarray}
H(A,\phi,\xi)=\frac{i\,(1+x/K)\exp\big(-i(x\xi/K + \nu\psi/Q_s)\big)}
{1-\exp\big(-2\pi i (x+\nu)\big)} \nonumber \\
\times\,\bar Y\Big(A\cos\big(\phi+\psi(A,\phi,\xi)\big)\Big).
 \quad
\end{eqnarray}
If $\,|\nu/Q_{s0}|<\sim1$, the function $H$ determined by Eq.~(5.11) satisfies the same condition as the function $H(\xi)$ by Eq.~(4.23): 
namely, it changes relatively little at a variation $\delta\xi\sim 2\pi$.
Therefore, result of calculations with Eqs.~(5.10-11) looks like Eq.~(4.25): 
\begin{eqnarray}
Y(A,\phi) \simeq i\,\big[H(A,\phi,2\pi K)-H(A,\phi,0)\big] .
\end{eqnarray}
It follows from Eq.~(5.8) and (5.9) that 
$\psi=0\;{\rm or}\;2\pi$ at $\xi=0\;{\rm or}\;2\pi K$.
Therefore, substituting here Eq.~(5.11) and taking into account that $x/K\ll1$ obtain at last
\begin{eqnarray}
Y(A,\phi) \simeq \bar Y(A\cos\phi)=\bar Y(\tau).
\end{eqnarray}
It means that the denominator zeros can be compensated, 
and undamped oscillations of the bunch are possible in the upper cluster.
However, an open question remains concerning a characteristics of the coherent bunch oscillations in the lower cluster, that is at $\nu\sim-\Delta Q$.


\section{Numerical solution (the upper cluster).}


A model of the bunch equation is developed in this section.
It is based on Eq.~(2.10) which, being multiplied by the distribution function $F$ and integrating with respect to in $u$ at $\zeta_\tau=0$ and $w=1$, results in 
\begin{eqnarray}
Q_{s0}^2\int \frac{\partial}{\partial\phi}\left(\frac{1}{\hat\nu}
\frac{\partial Y^{(+)}}{\partial \phi}\right)\,
F\,du + \nu\lambda(\tau)\bar Y(\tau) =  2q\lambda(\tau)
\int_\tau^1\bar Y(\tau')\,\lambda(\tau')\,d\tau' \quad
\end{eqnarray}
where $\lambda(\tau)$ is the bunch linear density, and Eqs.~(2.9) are taken into account.
The key point of the proposed model being is a usage of the relation 
$Y^+\simeq\bar Y$  which has been found for the upper cluster of the eigenmodes, see Eq.~(5.13).
Substituting it in Eq.~(5.14) and applying the equivalence  
$$
\frac{\partial}{\partial\phi}=-u\,\frac{\partial}
{\partial\tau}+\tau\,\frac{\partial}{\partial u}
$$
obtain
\begin{eqnarray}
Q_{s0}^2\left[U^2(\tau)\frac{d}{d\tau}
\left(\frac{1}{\hat\nu}\frac{d\bar Y}{d\tau}\right)
-\frac{\tau}{\hat\nu}\frac{d\bar Y}{d\tau}\right] + \nu\bar Y(\tau)
=2q\int_\tau^1 \bar Y(\tau')\,\lambda(\tau')\,d\tau' 
\end{eqnarray}
where
\vspace{-5mm}
\begin{eqnarray}
U^2(\tau)=\frac{1}{\lambda(\tau)}\int F(\tau,u)\,u^2\,du
\end{eqnarray}
Taking the distribution given by Eq.~(5.2) as an example, obtain
\begin{eqnarray}
F = \frac{2\alpha+1}{2\pi}(1-A^2)^{\alpha-1/2},\quad 
\lambda(\tau)=C_\alpha(1-\tau^2)^\alpha, \quad  
U^2(\tau)=\frac{1-\tau^2}{2(\alpha+1)}
\hspace{4mm}
\end{eqnarray}
with the normalizing coefficients $C_\alpha$ by Table~5.1.
Corresponding Eq.~(5.15) is
\begin{eqnarray}
\!\!\!\!\frac{1-\tau^2}{2(\alpha\!+\!1)}\,\!\frac{d}{d\tau}
\left(\frac{1}{\hat\nu}\frac{d\bar Y}{d\tau}\right)
-\frac{\tau}{\hat\nu}\frac{d\bar Y}{d\tau} 
+\frac{\nu\bar Y(\tau)}{Q_{s0}^2}
=\frac{2q C_\alpha}{Q_{s0}^2}\int_\tau^1 \bar Y(\tau')(1\!-\!\tau'^2)^\alpha\,d\tau' 
\end{eqnarray}
It coincides with Eq.~(4.14) at $\,\alpha=0\,$ because $\,\hat\nu\,$ is a constant  in the case. 
Also note that, at $q=0$ and $\Delta Q_0=0$, eigenfunctions of Eq.~(5.18) are 
the Gegenbauer polynomials $G_n^{\alpha+1/2}(\tau)$ with the eigentunes 
$\,\nu^2/Q_{s0}^2=n(n+2\alpha+1)/(2\alpha+2)$
\footnote{G. A. Corn and T. M. Corn, "Mathematical Handbook" 21.7-8, Dover Publications. Inc., Mineola, NY (1968).} .
In particular, they are the Legendre polynomials at $\alpha=0$.

The case $\,\alpha=1\,$ (parabolic bunch) is considered below as an example. 
After simple transformations, the equations are in the case:
\begin{subequations}
\begin{eqnarray}
\frac{1-\tau^2}{4}\,\!\left(\frac{\bar Y'(\tau)}{\hat\nu(\tau)}
\right)'-\frac{\tau\bar Y'(\tau)}{\hat\nu(\tau)}+\frac{\nu\bar Y(\tau)}{Q_{s0}^2} = 
\frac{3q}{2Q_{s0}^2}\int_{-1}^\tau\bar Y(\tau')(1\!-\!\tau'^2)\,d\tau',\quad 
\end{eqnarray}
\vspace{-7mm}
\begin{eqnarray}
\hat\nu(\tau) = \nu+\Delta Q_0(1-\tau^2), 
\end{eqnarray}
\vspace{-7mm}
\begin{eqnarray}
\bar Y'(-1) = -\frac{\nu^2\bar Y(-1)}{Q_{s0}^2}, 
\end{eqnarray}
\vspace{-7mm}
\begin{eqnarray}
\bar Y'(1) = \frac{\nu^2\bar Y(1)}{Q_{s0}^2}-\frac{3q}{2Q_{s0}^2}\int_{-1}^1\bar Y(\tau')(1\!-\!\tau'^2)\,d\tau'.
\end{eqnarray}
\end{subequations}
Solutions are found numerically at initial conditions 
$\bar Y(-1)=1,\;\bar Y'(-1)=-\nu^2/Q_{s0}^2$, 
by a selection of the parameter $\nu$ to provide the required boundary condition 
(5.19d) at $\tau=1$.  
Some results are represented in Figs.~5.1-5.3.
Two of them are obtained at $q=0$ and show the lower eigenfunctions and eigentunes of the bunch. 
The last plot demonstrates the tunes coalescence against the wake strength giving a possibility to find the TMCI threshold at different SC tune shift.
More details are placed in the figure captions where a noticeable similarities of this case
with the flat bunch case (Sec.~4. ) is noted.
\\

However, it should be kept in mind that these conclusions were obtained in the 
limit of $\Delta Q/Q_{s0}\rightarrow\infty$, and refer only to the upper 
cluster of the eigentunes.
In other (not considered) cases, significant differences may be expected due to the SC tune spread, which is out of the "flat" model.

\begin{figure}[p!]
\vspace{-22mm}
\centerline{\epsfig{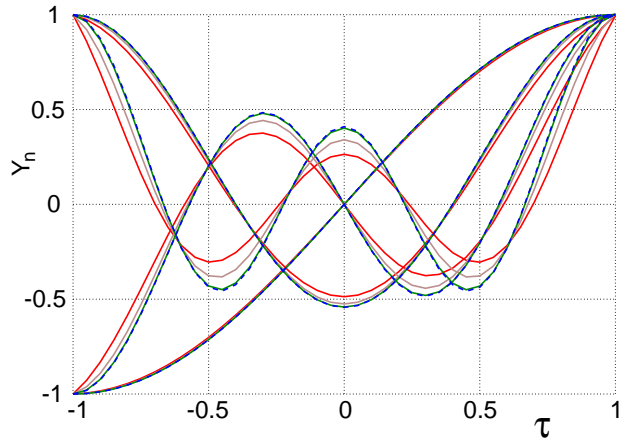}}
\caption{Eigenfunctions of the parabolic bunch $\bar Y_n$ without wake at $n=1,2,3,4$
(the upper cluster).
The function index matches the number of its zeros.
The trivial solution $\bar Y_0=1$ is not shown.
The colors indicate the SC tune shift value:
Red, Green, Brown, Blue(dashed) $\Rightarrow \Delta Q_0= 3,\,5,\,10,\,30$.
The curves are indistinguishable in practice at bigger $\Delta Q_0$.
There is an essential similarity with the eigenfunctions of a flat bunch (Legendre polynomials).}
\end{figure}
\begin{figure}[b!]
\centerline{\epsfig{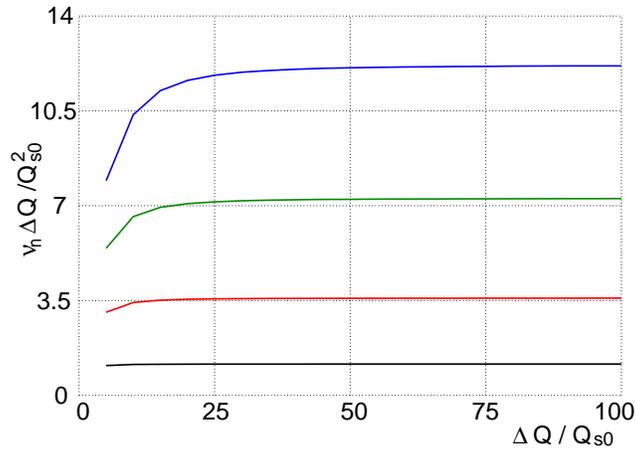}}
\caption{Eigentunes of the parabolic bunch $\nu_{nn},\, n=1,\,2,\,3,\,4\,$ without wake (the upper cluster). 
The indexes are matched with Figs.~5.1 being: $\,n=1,\,2,\,3,\,4\,$ from the bottom curve to the top one. 
The ultimate values are: $\,\nu_{nn}\Delta Q_0/Q_{s0}^2=kn(n+1),\,k\simeq 0.6$
($k=1\,$ for the flat bunch, see Eq.~(4.19)).} 
\end{figure}
\begin{figure}[b!]
\centerline{\epsfig{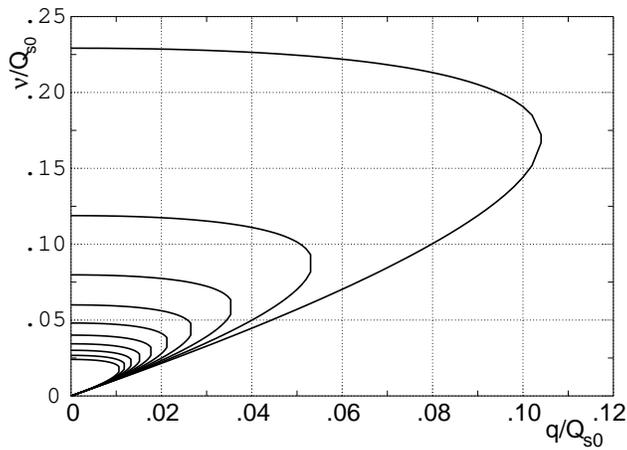}}
\caption{A coalescence of the eigentunes $\,\nu_{00}\,$ and $\,\nu_{11}\,$ of the parabolic bunch against the wake strength at different value of the SC tune shift 
(the zero mode was not shown in Figs.~5.1-2 because of the triviality: 
$\bar Y_0=1,\, \nu_{0,0}=0$. 
The curves in the figure are obtained at $\,\Delta Q_0/Q_{s0}$ changing from 5
(the far right line) to 50 with step 5.
The coalescence happens at $\,q\simeq 0.53\, Q_{s0}^2/\Delta Q_0\,$ what is very close to the flat bunch TMCI threshold, which is: $\,q\simeq 0.475\, Q_{s0}^2/\Delta Q_0\,$ 
at $\,\Delta Q/Q_{s0}\gg 1$, according to Eq.~(4.40).}
\end{figure}

\end{document}